\documentclass[referee]{aa} 
\usepackage{natbib}
\usepackage[english]{babel}

\usepackage{graphicx, subfigure}
\usepackage{psfig, epsfig}
\psfigurepath{./Figures/}

\usepackage{amsfonts, amsmath, amssymb}

\vspace{1cm}

\newcommand{\bitem}{\begin{itemize}}
\newcommand{\eitem}{\end{itemize}}
\newcommand{\benum}{\begin{enumerate}}
\newcommand{\eenum}{\end{enumerate}}
\newcommand{\beq}{\begin{equation}}
\newcommand{\eeq}{\end{equation}}

\newcommand{\unit}{\,\rm}

\begin{document}

\title{Cosmology with wide-field SZ cluster surveys: Selection and Systematic Effects}
       
\author{J.B. Juin,  D. Yvon, A. R\'efr\'egier \and C. Y\`eche}
\institute{CEA/DSM/DAPNIA, CE Saclay, Bat 141, F-91191 Gif-sur-Yvette Cedex, France}

\abstract{
The cosmological potential of large-scale structure observations for cosmology have been extensively discussed in the litterature. In particular, it has recently been shown how Sunyaev-Zel'dovich (SZ) cluster surveys can be used to constrain dark energy parameters.
In this paper, we study whether selection and systematics effects will limit future wide-field SZ surveys
from achieving their cosmological potential. For this purpose, we use a sky simulation and an SZ-cluster detection software presented in \citep{PiresJuin2005}, using the future Olimpo, APEX and Planck surveys as a concrete examples. We show that the SZ-cluster selection function and contamination of SZ-cluster catalogues are more complex than is usually assumed. In particular, the simulated field-to-field detected cluster counts is a factor 3 larger than the expected Poisson fluctuations. We also study
the impact of missing redshift information and of the uncertainty of the scaling relations for
low mass clusters. We quantify, through hypothesis tests, how near-future SZ experiments can be used to discriminate between different structure formation models. Using a maximum likelihood approach, we then study the impact of these systematics on the joint measurement of cosmological models and of cluster scaling relations. 
}

\maketitle 
\markboth{SZ clusters selection function}{}
\keywords{Cosmology, Clusters of galaxies, Sunyaev-Zel'dovitch effect, Data Analysis}

\section{Introduction}
In the next few years, a new generation of dedicated instruments based on large-array bolometer cameras (APEX\footnote{http://bolo.berkeley.edu/apexsz}, ACT\footnote{http://www.hep.upenn.edu/$\sim$angelica/act/act.html}, BOLOCAM\footnote{http://astro.caltech.edu/$\sim$lgg/bolocam\_front.htm}, OLIMPO \citep{exp:Olimpo}, and improved interferometers (AMI\footnote{http://www.mrao.cam.ac.uk/telescopes/ami/index.html}, AMiBA\footnote{http://www.asiaa.sinica.edu.tw/amiba}, SZA\footnote{http://astro.uchicago.edu/sze}) will provide large amount of information on cosmic structure formation and evolution, and thus on cosmological models. The Planck satellite \citep{exp:Planck}, to be launched in 2007, will provide a full-sky catalogue of galaxy clusters detected by their Sunyaev-Zel'dovich (SZ) signal \citep{Sunyaev1970, Sunyaev1972}. The potential of SZ observations results from the properties of the SZ effect: the lack of surface dimming and the "clean" measurement of thermal energy of the cluster gas, should afford a measure of the cluster mass function up to high redshift, with reduced systematics when combined with X-Ray observations and weak lensing surveys. The distribution of cluster abundance with redshift is sensitive to the cosmological parameters $\sigma_8$ and $\Omega_M$, and also to a lower extend to $\Omega_{\Lambda}$ and the dark energy equation-of-state \citep{Barbosa1996, Oukbir1997, Haiman2001}. \cite{Battye2003} studied the dependence of these cosmological constraints on large scale structure formation and gas cluster physics models. \cite{Melin2005} presented a first study of the selection function of large SZ-cluster survey. 

In this paper, we use  simulations of the sky and of an SZ experiment, along with a recent cluster detection pipeline presented in \cite{PiresJuin2005}, to simulate future large-array bolometer observations and cluster detections. We first discuss photometric issues, and then present a detected cluster catalogue, with its selection function, and purity curves. Those are found to be complex. In particular, the contamination of the cluster catalogue is quantified as a function of cluster brightness. We also compute the count variance in the observed catalogues.
We then assume that the observations and cluster detection can be statistically described by an observation model. This observation model allows us to transform a semi-analytic cosmology-motivated cluster count functions, $\frac{dn}{dz \, dY(z,Y)}$, into a set of probability density functions (pdf) of the detected cluster observed parameters $N_{Obs}, Y_{Obs}, z_{Obs}$ and of the contaminants $N_{Cont}$ and $Y_{Cont}$, where $Y$ is the Compton parameter integrated over the cluster angular size, and $z$ the redshift. This observation model is found to be accurate enough given the statistics of the upcoming SZ cluster surveys and to be computationally very effective. 
Based on this model, we then discuss our ability to constraint cosmology assumptions and parameters. We show, using an hypothesis test method, how future SZ surveys will make it possible to distinuigh between several mass functions. We quantify the constraint that such an experiment would place on the effective ``heating" parameter $T_*$, using the cosmological parameters values measured by WMAP to break degeneracies.
We conclude by showing how any conclusions on cosmological parameters are sensitive to inaccuracies in the observation model.


Future large-array bolometer surveys share some common features. They observe the sky in several frequency bands to facilitate astronomical source separation. They use large bolometer matrices to maximise redundancy of observation on the sky and speed up field coverage. They are ambitious in terms of mission length, given the technology. In this paper, we mostly use the Olimpo project as
a concrete example of an upcoming bolometer-based large SZ survey.


\section{Montecarlo simulations}
	
In the following, we use the sky simulation software, the instrument model and the cluster detection pipeline described in \citep{PiresJuin2005}. We briefly summarise it here for convenience. We simulated four astrophysical contributions to the sky map: primordial CMB anisotropies (excluding the dipole), bright infrared galaxies as observed by SCUBA \citep{Borys2003}, the infrared emission of the Galaxy and SZ-clusters. All simulations uses a cosmological model with parameters shown in table \ref{CosModParam}.

\begin{figure}[ht]
\begin{minipage}[l]{0.5\linewidth}
\centering
\includegraphics[width=6cm]{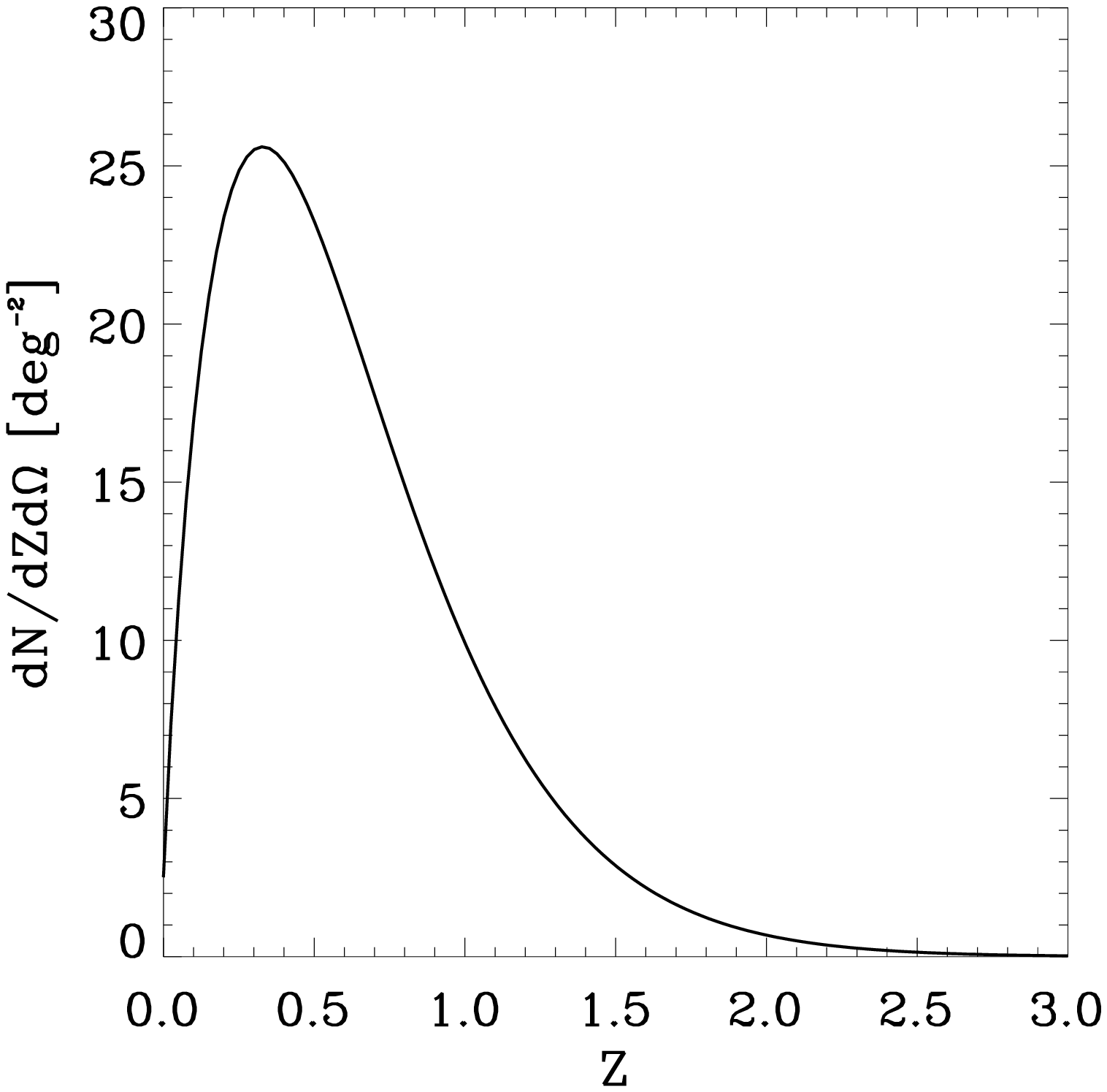}
\end{minipage} 
\hfill
\begin{minipage}[r]{0.5\linewidth}
\centering
\includegraphics[width=6cm]{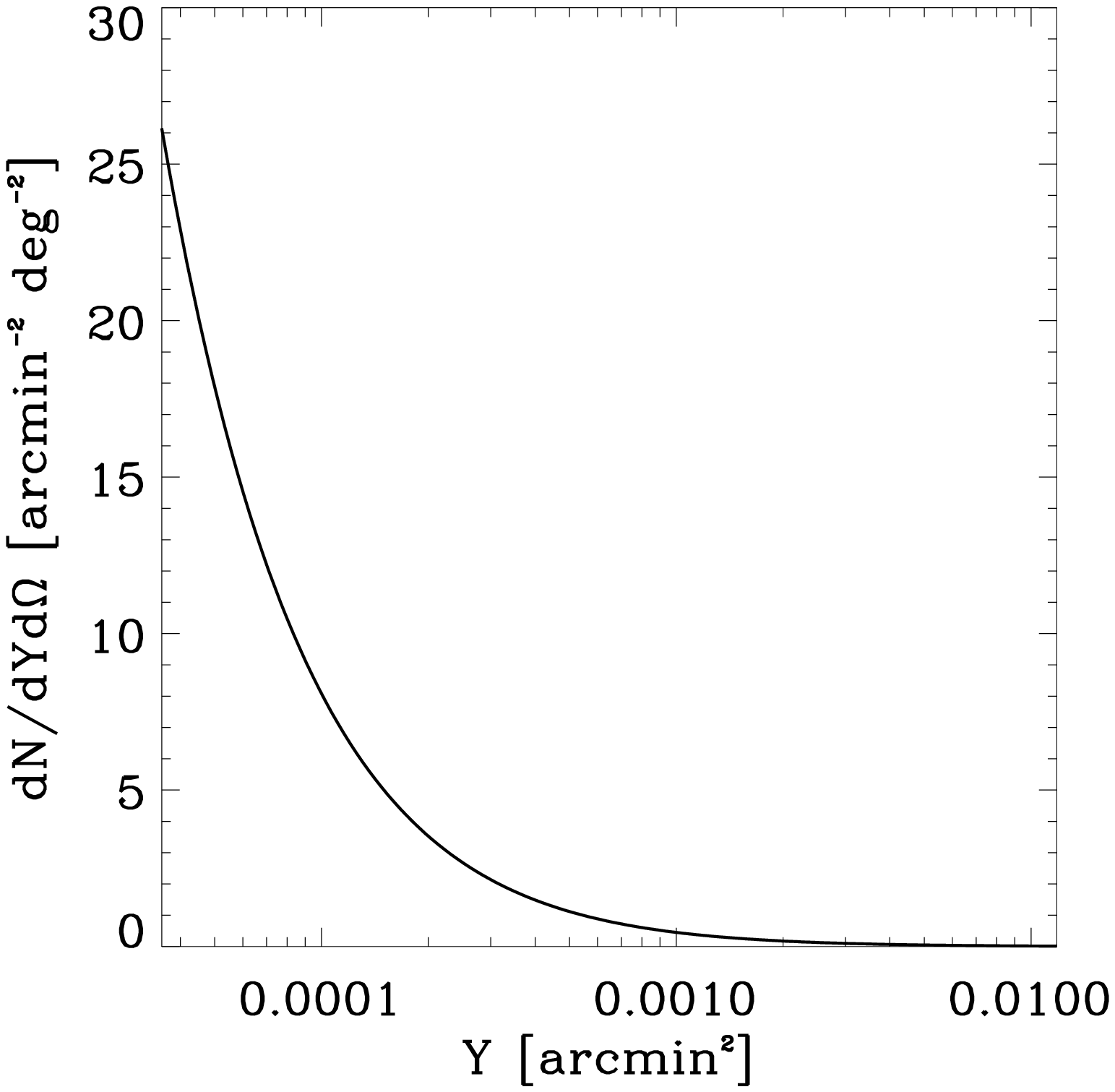}
\end{minipage}
\hfill
\caption[]
{Generated cluster distribution as a function of redshift and integrated Compton flux.}
\label{TheoClusDistrib}
\end{figure}

\begin{table}
\centering
\begin{tabular}{|c|c|c|c|c|c|c||c|c|c|}
\hline
$\Omega_{tot}$ & $\Omega_b$ &  $\Omega_{\Lambda}$  & $\Omega_{DM}$ & $h$ & $n_s$ & $\sigma_8$  & $\rm f_{mass}$ & $T_\star$ & $f_g$  \\
\hline
$1$ & 0.05 & 0.7  & 0.25 & 0.7 & 1 & 0.85 & S. \& T. & 1.9 & $0.9 \, \Omega_b / \Omega_m$ \\
\hline
\end{tabular}
\caption{Cosmological and gas physics parameters used in the simulations. Densities relative to critical density are labelled $\Omega$.
$\Omega_{tot}$ is the density of universe, all components included, $\Omega_b$ is the baryon density, $\Omega_{\Lambda}$ the vacuum energy density, $\Omega_{DM}$ the dark matter density. $h$ is the reduced Hubble constant, $n_s$ the spectral index of primordial density power spectrum, $\sigma_8$ the rms density fluctuations in spheres of $8 \unit MPc$ diameter, $f_{Mass}$ the mass function used in cluster abundance computations, $T_{\star}$ the cluster mass-temperature normalisation factor, and  $f_g$ the cluster gas mass fraction. The double vertical line destinguishes the primordial cosmological parameters, from the ingredients of the structure formation semi-analytic model. }
\label{CosModParam}
\end{table}
Figure \ref{TheoClusDistrib} shows the distribution of the generated cluster as a function of redshift and integrated Compton flux.

The frequency dependence of the bright infrared sources and of galactic dust are described by a grey-body spectrum. The spectral index of bright infrared sources is generated randomly for each sources between 1.5 and 2. Table \ref{ExpOlimpo} provide the noise level and the FWHM of antenna lobe at each frequencies, which is assumed be Gaussian. Bandwidth filters are assumed have a top hat response.

%
\begin{figure}[ht]
\begin{minipage}[l]{.15\linewidth}
\centering
\includegraphics[width=4.5cm]{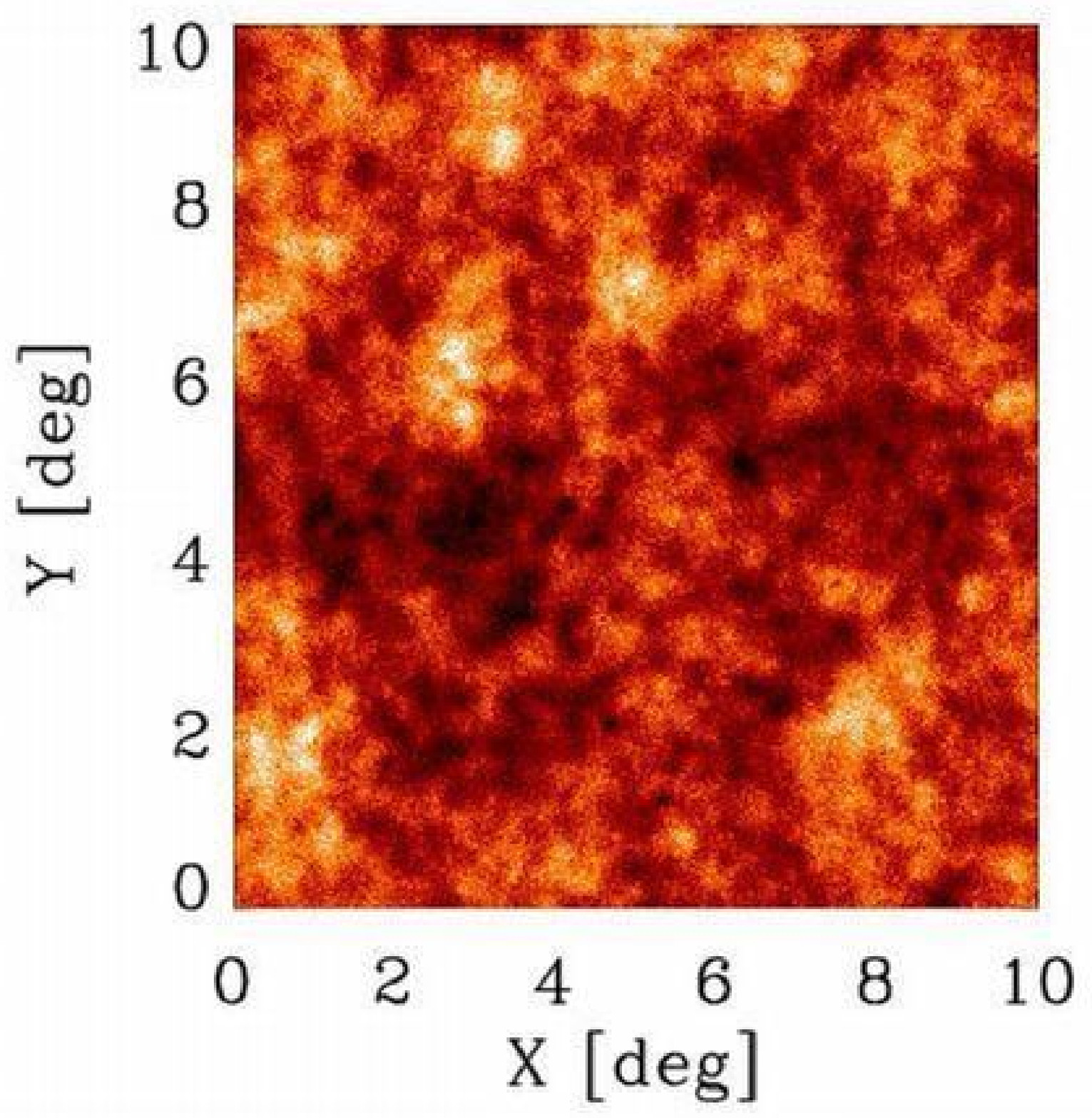}
\end{minipage} 
\hfill
\begin{minipage}[r]{.15\linewidth}
\centering
\includegraphics[width=4.5cm]{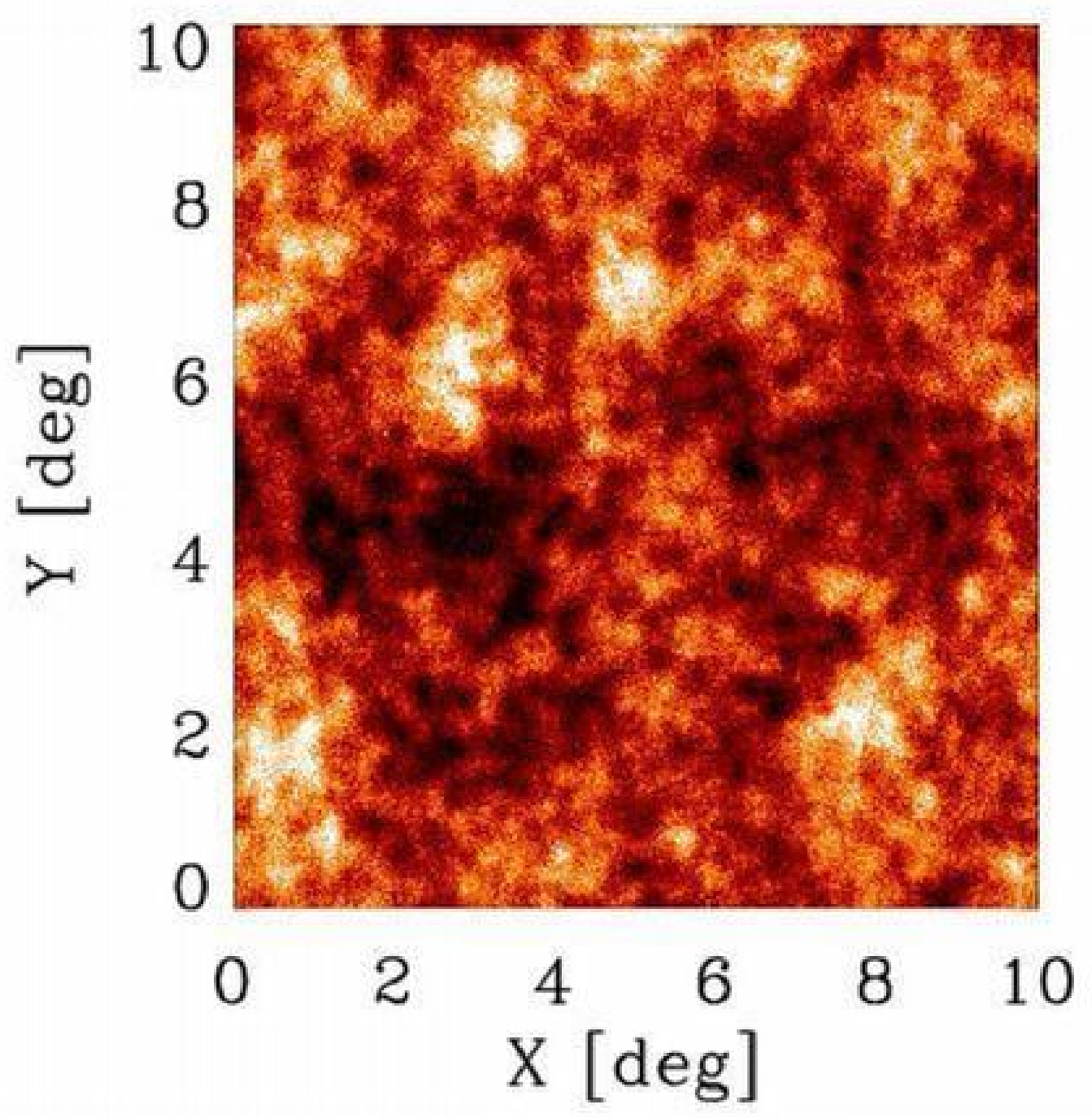}
\end{minipage}
\hfill
\begin{minipage}[c]{.15\linewidth}
\centering
\includegraphics[width=4.5cm]{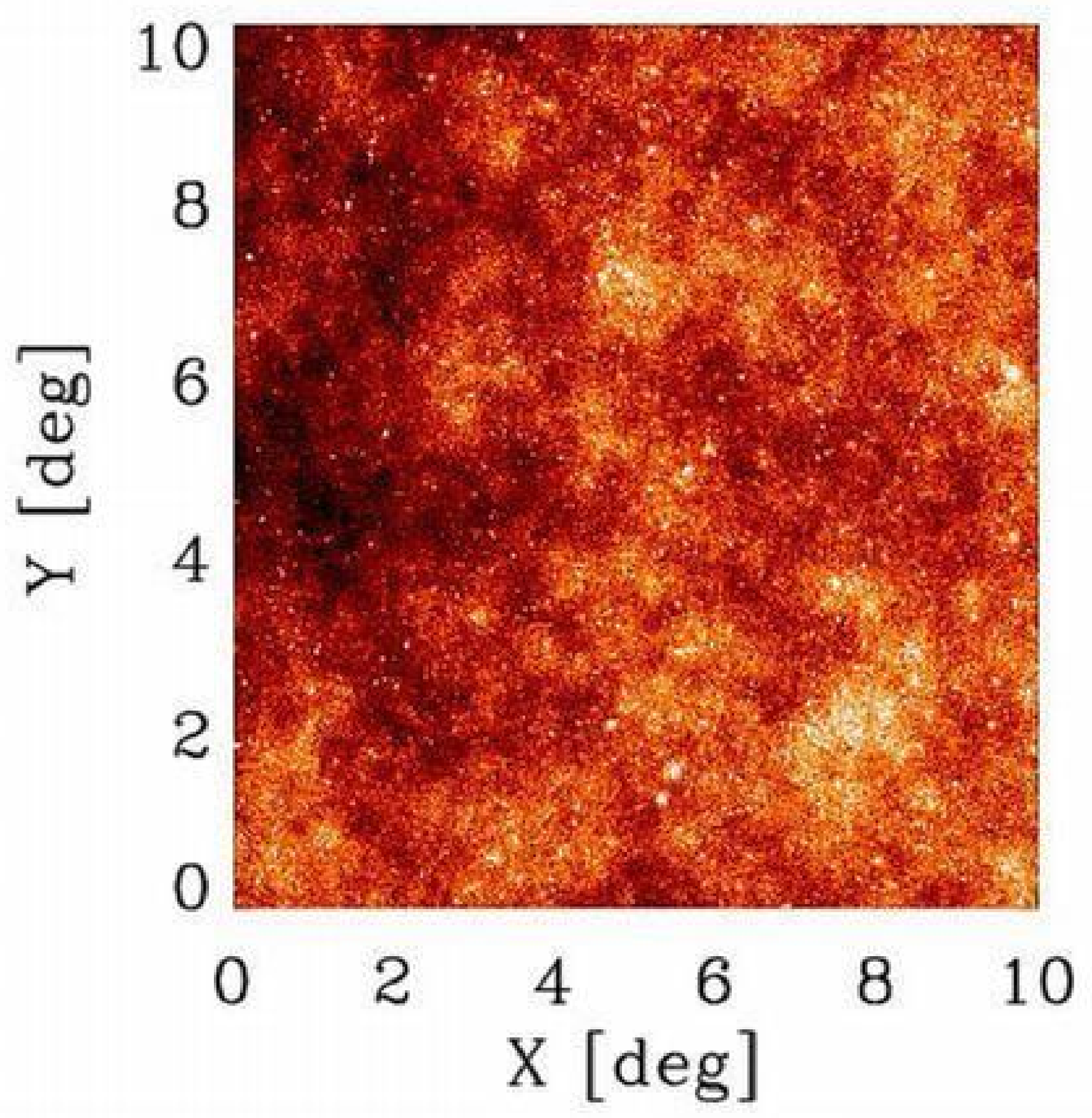}
\end{minipage}
\hfill
\begin{minipage}[r]{.15\linewidth}
\centering
\includegraphics[width=4.5cm]{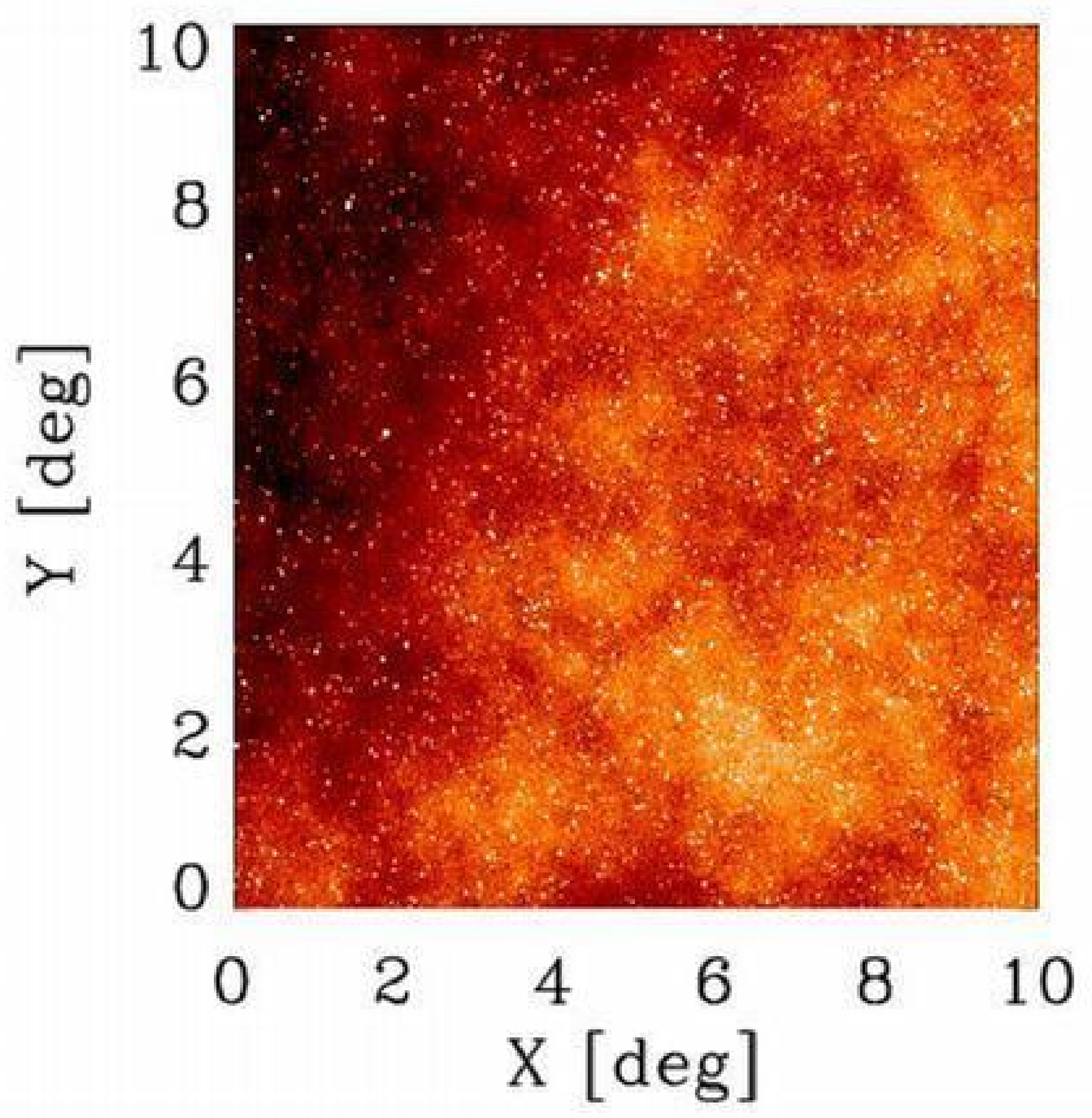}
\end{minipage}
\caption[Observation simulations]
{Simulations of sky maps, as observed by a large-array bolometer experiment. For these simulations, we used the Olimpo experiment model. From Left to right, 143, 217, 385, and 600 GHz bands are shown. In the two lower frequencies band, CMB primordial anisotropies are the dominant features. At  higher frequencies bands, bright Infrared galaxies and Galactic dust become dominant. The  SZ cluster signal is sub-dominant at all frequencies.}
\label{CartesSimulees}
\end{figure}

Figure \ref{CartesSimulees} shows the `observed' maps, simulated using the Olimpo parameters.
We then apply an Independent Component Analysis method named JADE \citep{Cardoso1999} on our map, after a wavelet transform. JADE separates the SZ signal from the other astrophysical sources effectively, as long as the noise level is kept low enough, and provides a noisy SZ map. We apply then a Multiscale Entropy False Discovery Rate filter \citep{Starck2005} on the map and detect sources using the SExtractor software \citep{BertinArnouts1996SExtractor}.
Detected sources can be reliably associated with simulated clusters and are labelled as 'true' clusters. This allows us to compute the selection function and the photometric accuracy of our simulated observations. Detections that are not associated with simulated clusters are identified as contaminants. We can then calculate the purity of our recovered sample and the brightness distribution of the contaminants.

\begin{table}
\centering
\begin{tabular}{|c||c|c|c|c|}
\hline
Observation bands $\nu$ [$\unit GHz$] & 143 & 217 & 385 & 600 \\
\hline
\hline
Bandwidth $\Delta \nu$  [$\unit GHz$] & 30 & 30 & 30 & 30 \\
\hline
Bolometer Number & 19 & 37 & 37 & 37 \\
\hline
Lobe Width, FWHM [$\unit arcmin$] & 3 & 2 & 2 & 2 \\
\hline
Noise level [$\unit \mu K_{CMB} s^{1/2}$] & 150 & 200 & 500 & 5000 \\
\hline
\end{tabular}
\caption{Foreseen experimental features of the Olimpo balloon bolometer project.}
\label{ExpOlimpo}
\end{table}

\subsection{Photometry} \label{PhotErrorsChap}
\begin{figure}[ht]
\begin{minipage}[l]{0.5\linewidth}
\centering
\includegraphics[width=8cm]{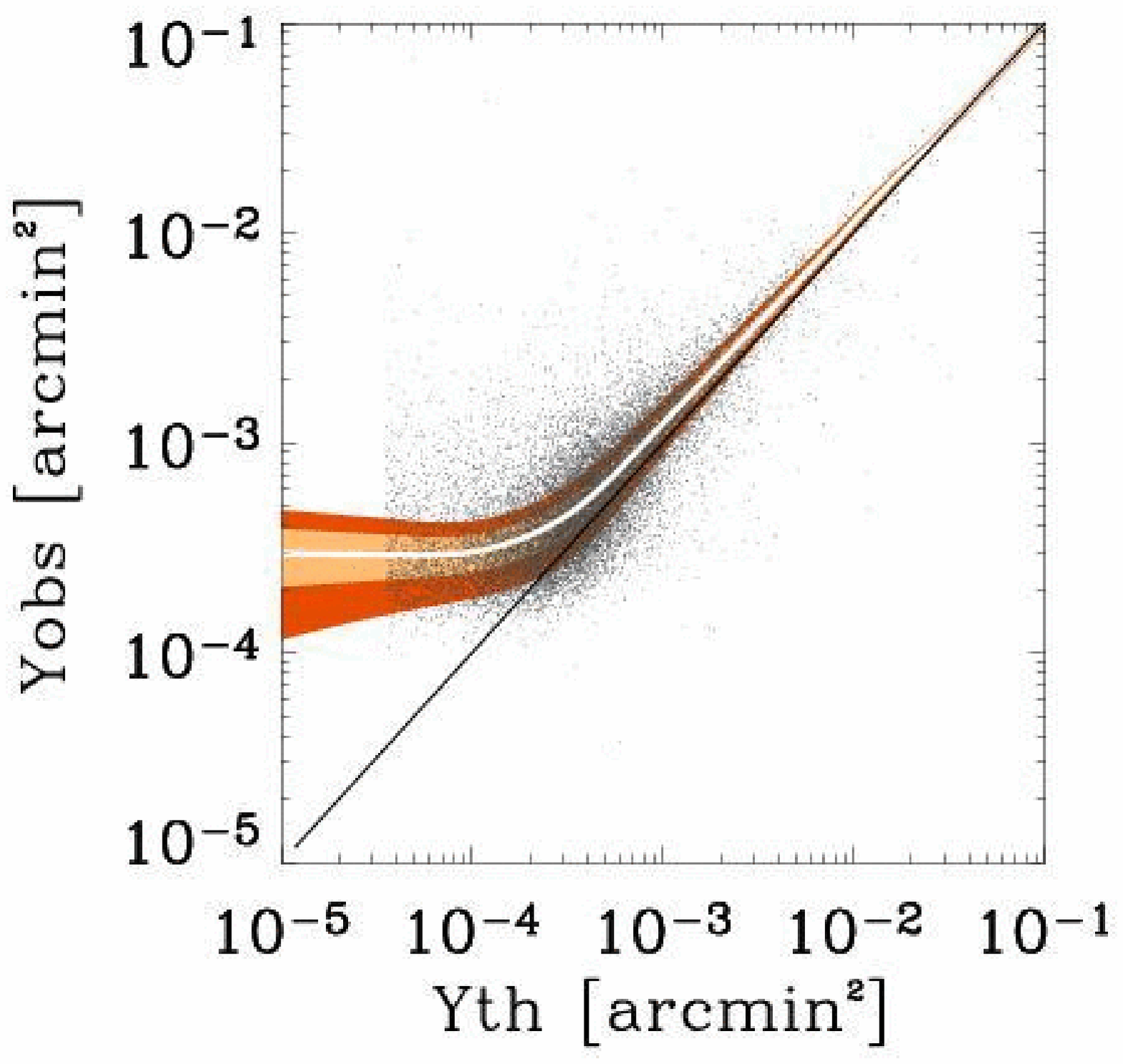}
\end{minipage} 
\hfill
\begin{minipage}[r]{0.5\linewidth}
\centering
\includegraphics[width=8cm]{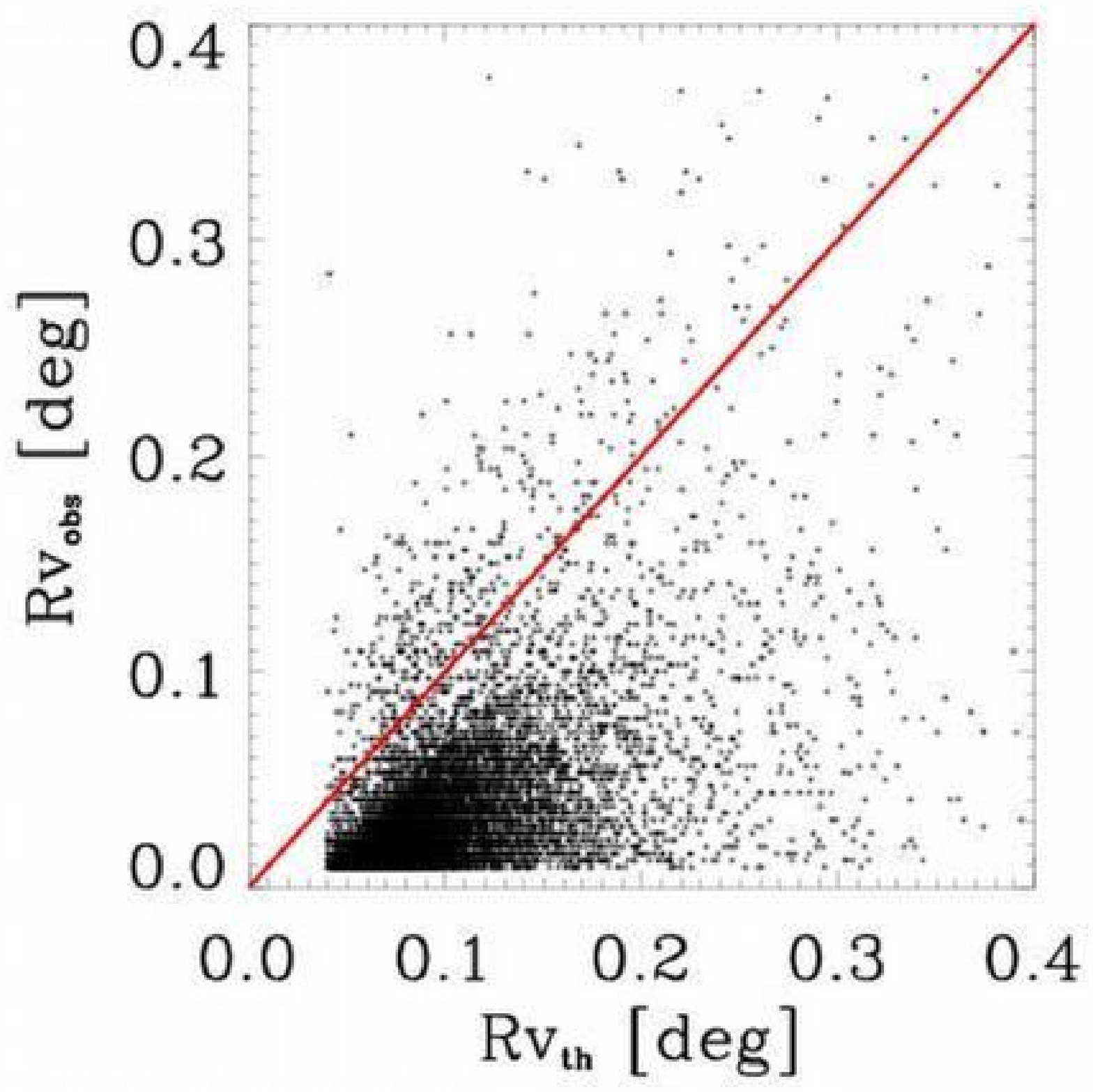}
\end{minipage} 
\caption[]{
\textbf{Left:} cluster reconstructed flux versus the true simulated flux, and our photometry model contours. 20 cumulative Monte-Carlo simulations where used for this plot.\\
\textbf{Right:} SZ cluster reconstructed virial size versus true simulated virial size. No correlation is seen.}
\label{PhotoFluxBiasVirSiz}
\end{figure}

Selecting the sources associated with simulated clusters, we plot in figure \ref{PhotoFluxBiasVirSiz} the observed cluster flux $Y_{Obs}$ versus the true simulated flux $Y_{th}$, and derive our photometry model, i.e. the probability density function pdf($Y_{Obs}|Y_{th}$). The observed flux is strongly overestimated at low brightness due to the Malmquist-Eddington bias \citep{Malmquist}. Our first attempt for a statistical model reproduces very well the simulated photometric behaviour, except for the (small) non-Gaussian tails. 

\subsection{Cluster Size reconstruction}
One way to infer the redshift of a cluster would be to measure its virial radius. Figure \ref{PhotoFluxBiasVirSiz} plots the reconstructed clusters virial radius versus their true (simulated) virial radius.
One can hardly see a significant correlation between the simulated virial radius and the observed virial radius. This is the reason why we decided to neglect this information in the following work.

\begin{figure}[htp]
	\begin{minipage}[l]{0.30\linewidth}
	\centering
	\includegraphics[width=5.5cm]{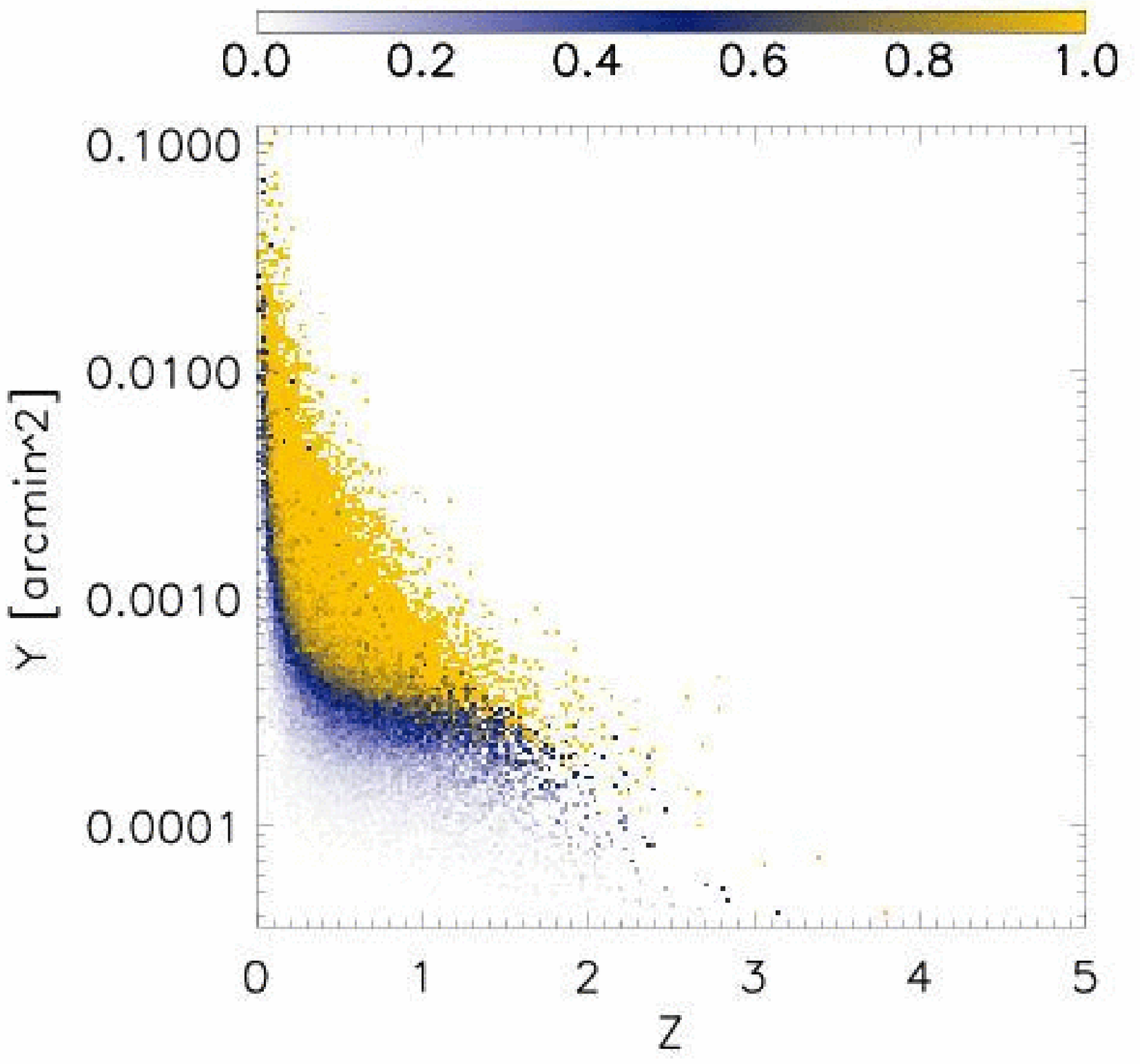}
	\end{minipage} 
	\hfill
	\begin{minipage}[c]{0.30\linewidth}
	\centering
	\includegraphics[width=5.5cm]{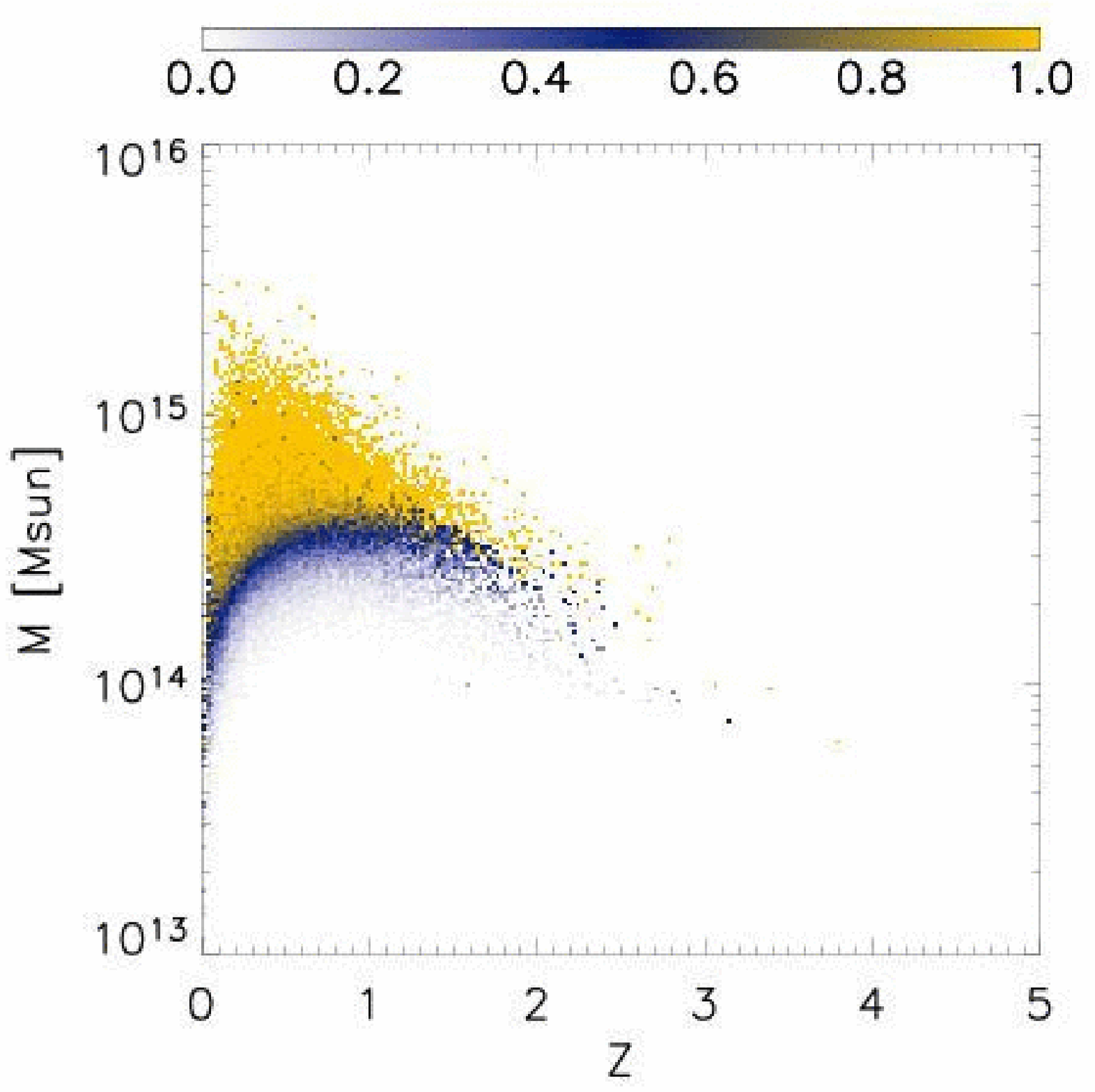}
	\end{minipage} 
	\hfill
	\begin{minipage}[r]{0.30\linewidth}
	\centering
	\includegraphics[width=5.5cm]{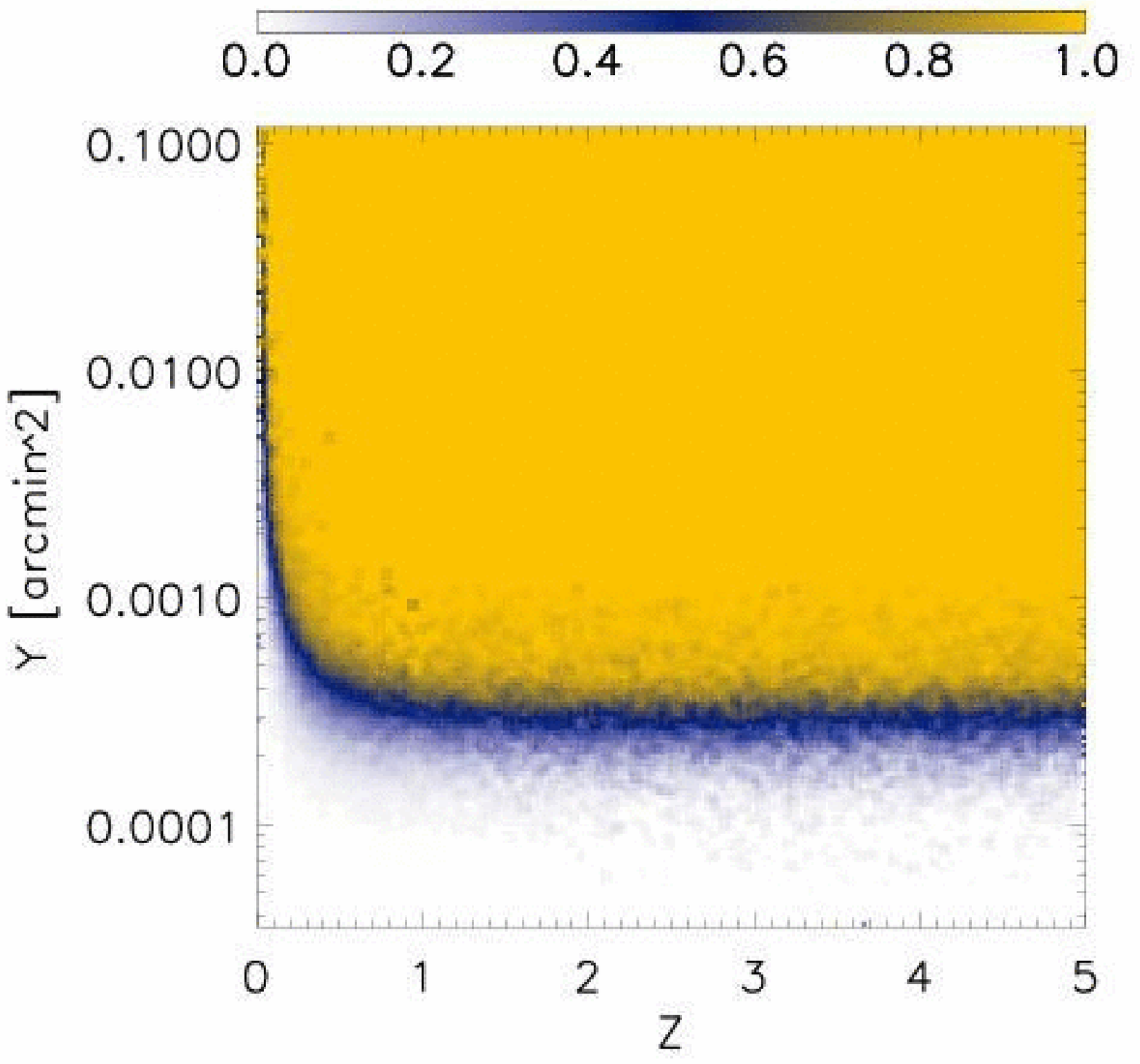}
	\end{minipage} 
	\caption[]
{Completness as a function of redshift, flux (\textbf{left}) and mass (\textbf{middle}), as simulated from a semi-analytic large scale structure and cosmology model. We used design parameters of the Olimpo project to model observation performance. \textbf{Right:} modelled selection function after extended simulations.}
	\label{ComplMonteCarl}
\end{figure}

\subsection{Completeness}\label{CMonteCarl}
From the true cluster catalogue, we computed the cluster detection probability as a function of cluster integrated fluxes, redshifts and masses. Figures \ref{ComplMonteCarl} left and middle, show the results. We see that a selection function can not be taken as a simple cut in total flux, nor in mass.
We also notice that clusters at large redshift are detected, even though very few are predicted by the
cosmological model. To quantify the selection function  at high redshift, we therefore introduced ``by hand" in our simulated map, 10 \% additional high-z clusters, randomly generated in the guessed $Y$-threshold area. We averaged 100 Monte-Carlos and we obtained the completness map plotted at figure \ref{ComplMonteCarl} right.
The selection function reduces to a simple $Y$ sensitivity curve at large redshift, when cluster sizes become smaller than angular resolution. But at redshift below $1$,  where we expect to detect most of the clusters, the completness curve is strongly distorted toward high Compton flux.
For convenience, we provide in Annex \ref{SelFuncAnnex} the tabulated values of Olimpo selection function versus cluster mass and redshift.

\begin{figure}[htb]
	\begin{minipage}[r]{.5\linewidth}
	\centering
	\includegraphics[width=7cm]{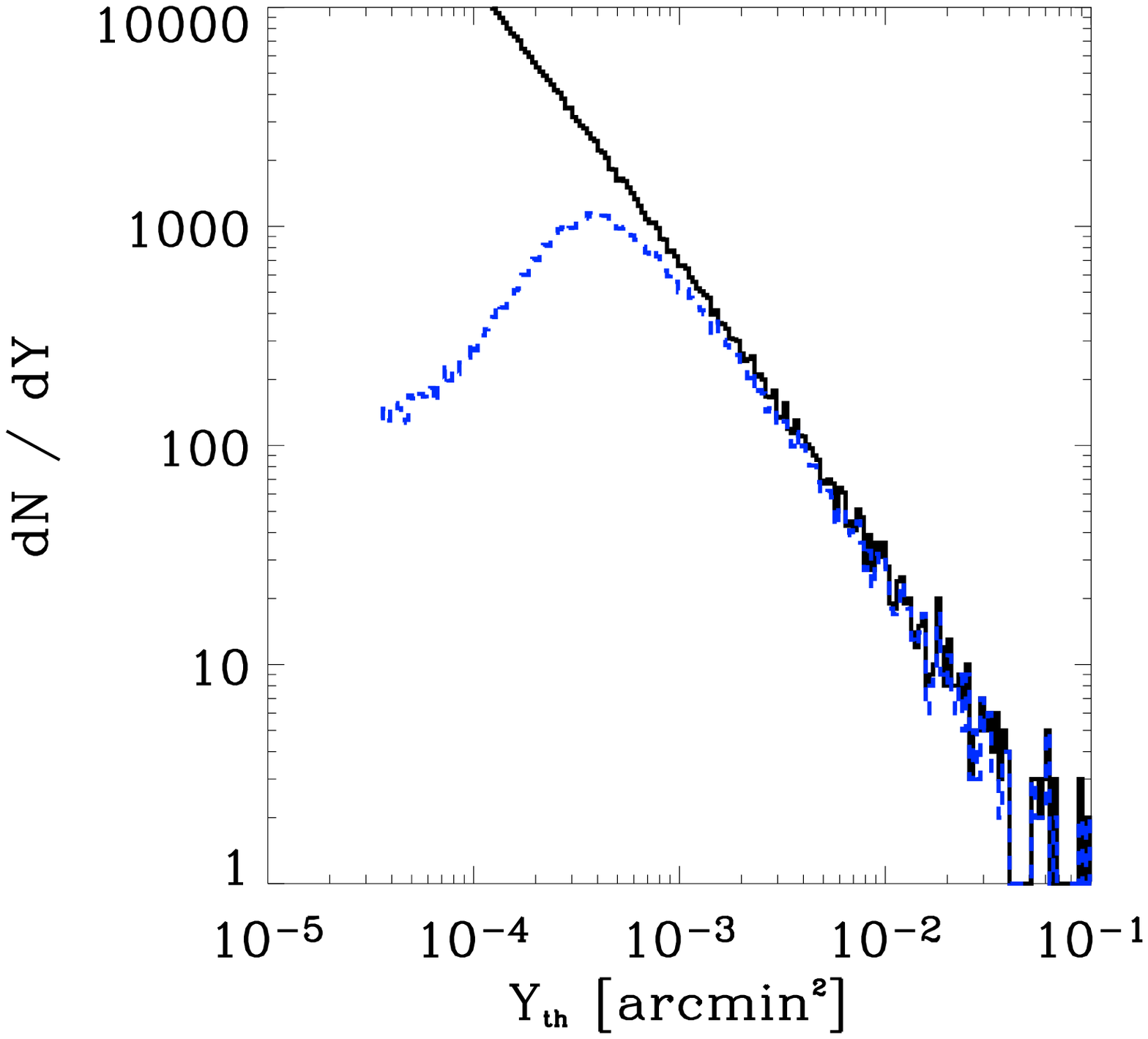}
	\end{minipage}
	\hfill
	\begin{minipage}[r]{.5\linewidth}
	\centering
	\includegraphics[width=7cm]{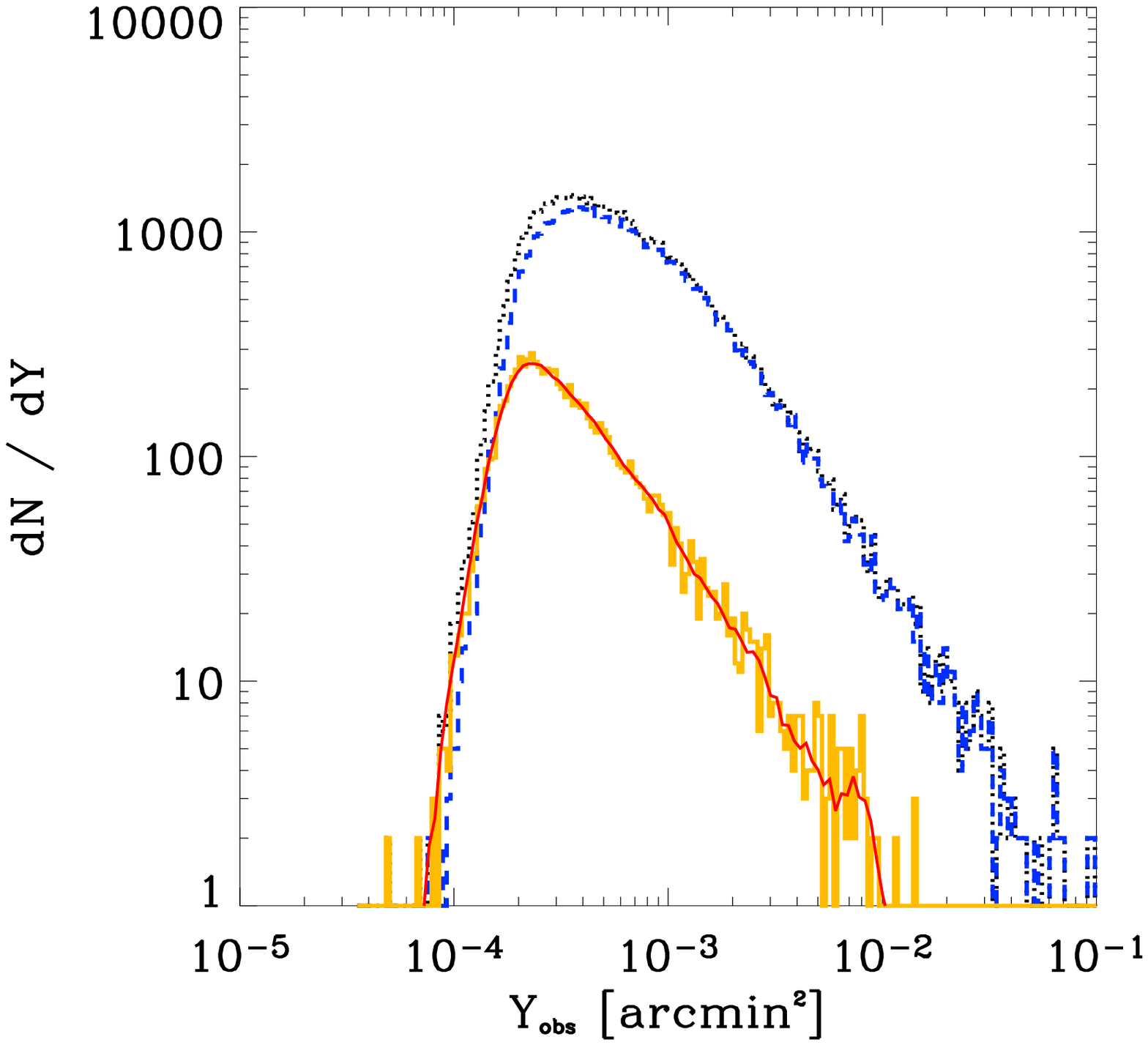}
	\end{minipage}
	\caption[]
	{We ran 100 montecarlo on 400 $deg^2$. \textbf{Left:} the black curve is the histogram of generated cluster flux, compared to the the blue histogram of true cluster detection.
\textbf{Right:} the blue histogram is the true cluster observed flux. The flux distribution of the contamination is plotted in orange. The red curve is our modelled flux distribution of contamination.}
	\label{DistribVermin}
\end{figure}

\subsection{Purity and contamination} \label{VerminChap}
Future SZ-cluster experiments won't be able to easily sort the contamination from the true clusters. Our evaluation of the observed flux distribution of contaminants is done by selecting sources that are not associated with simulated clusters, and is shown at figure \ref{DistribVermin}. The contamination histogram provides the red curve which we use as our modelled flux distribution of contamination. An integration over histograms shown in igure \ref{DistribVermin} lead to a sample purity value of 91\%, tuned by choosing the detection threshold.

\subsection{Sources counts}\label{SrcCounts}
\begin{figure}[hpt]
\begin{minipage}[l]{.3\linewidth}
\centering
\includegraphics[width=5.5cm]{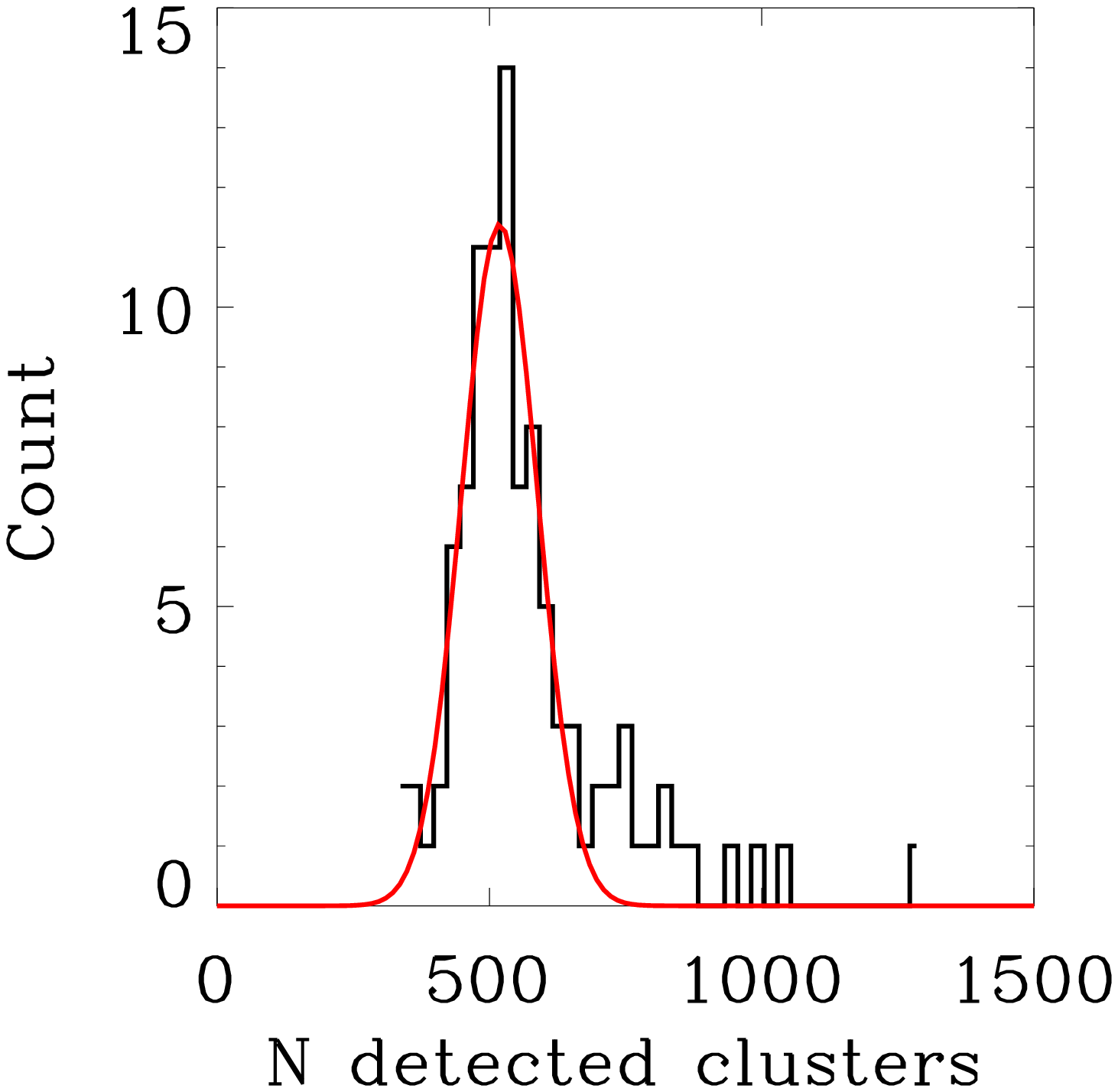}
\end{minipage}
\hfill
\begin{minipage}[c]{.3\linewidth}
\centering
\includegraphics[width=5.5cm]{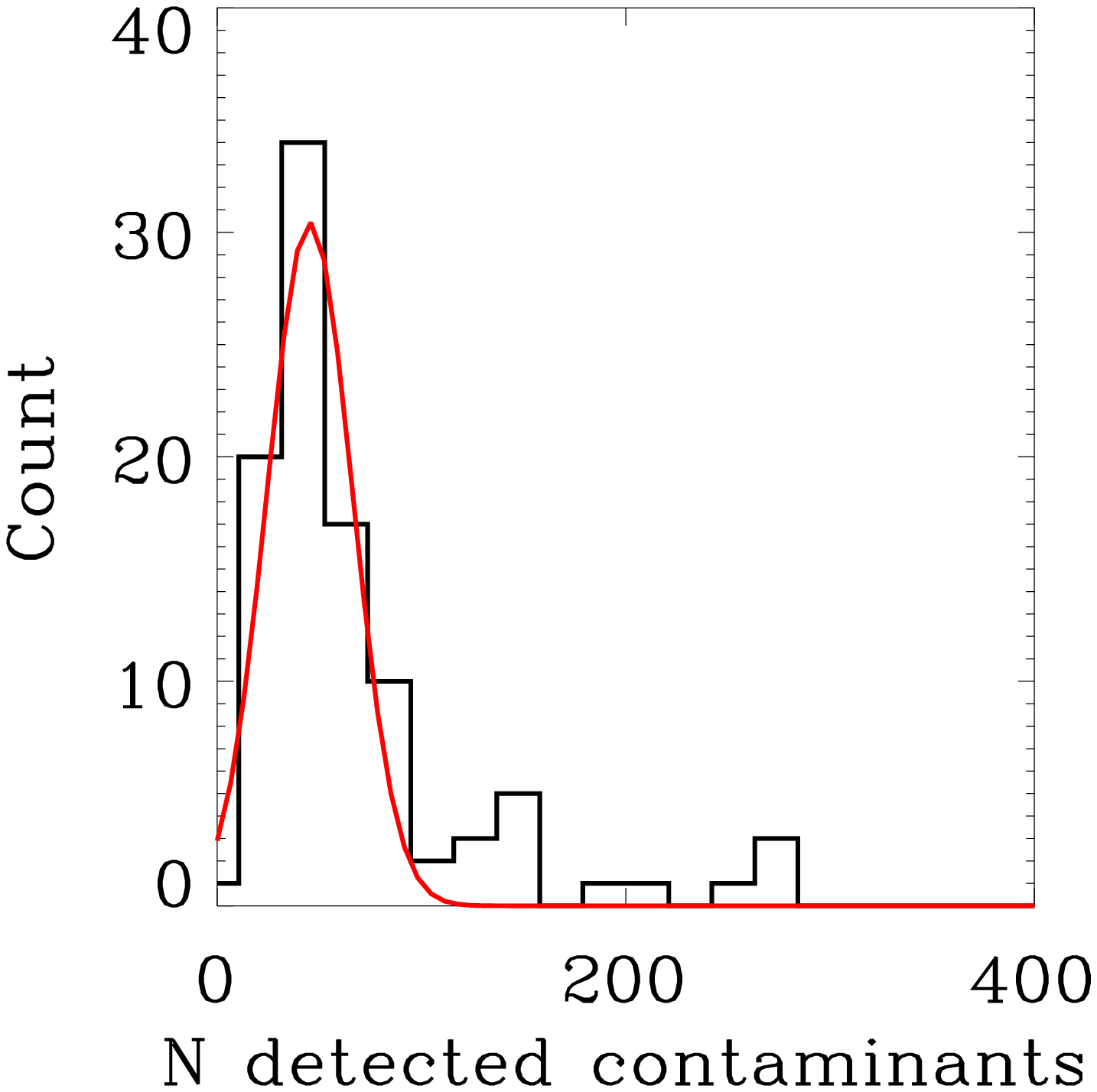}
\end{minipage}
\hfill
\begin{minipage}[r]{.3\linewidth}
\centering
\includegraphics[width=5.5cm]{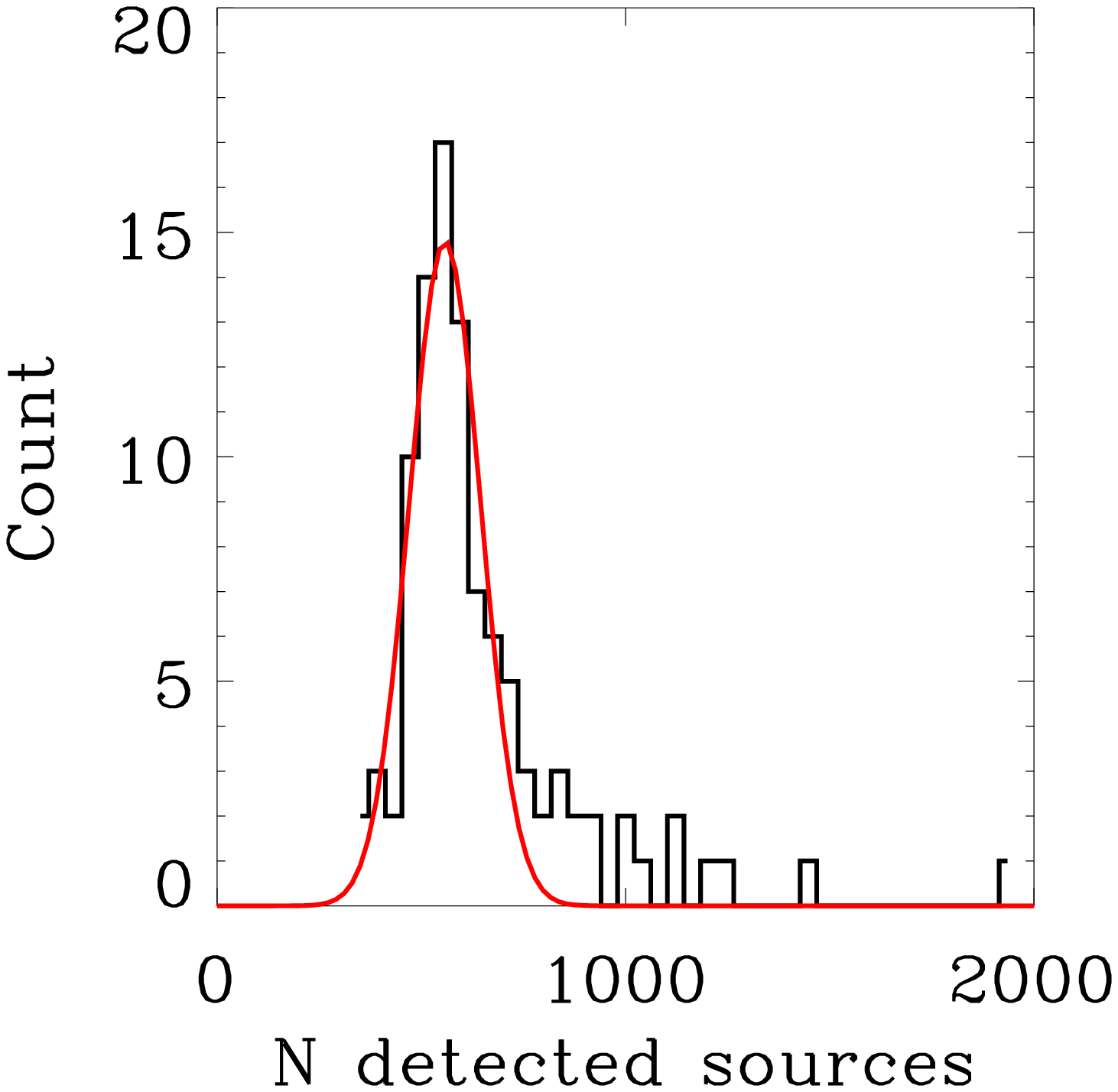}
\end{minipage}
\caption[]
{\textbf{From left to right:} true clusters, contamination and sources counts histograms for 100 simulations. Red curve fits are used in the following observations' model.}
\label{ModelNDetection}
\end{figure}

Cluster counts provides powerful information for large scale structure physics and cosmology. If the counts are dominated by field-to-field variations, one would expect them to follow a Poisson distribution. Figure \ref{ModelNDetection} shows the histograms of cluster, contamination and source counts for 100 simulated fields. We notice that all the histograms show non-Gaussian tails. A closer look shows that this happen when the confusion of few bright clusters, bias our SZ-map normalisation method and lead to a low value for the cluster detection threshold. As a result, the contamination and thus the total cluster counts increases. This simulation-only artifact will be solved in the future development of the detection
pipeline.

We also notice that the peak of the distributions  are well fitted by Gaussian curves. By computing the peak FWHM, we note a factor of 3 excess widths, relative to Poisson's distribution expectation. The origin of this excess is not yet settled. Possible explanations are other unidentified systematic effect in the cluster detection pipeline, or confusion due to the few large clusters, that mask smaller (but above threshold) cluster and thus induces lower statistic processes in the count variance. 
If it can not be avoided, cosmological constraints deduced from future observations will therefore have to take this effect into account. In the following we will use fits to the cluster and contamination counts (red curves in figure \ref{ModelNDetection}) to construct our observation model.

%

%
%


%
%
	
\section{Observation model} \label{StatModelObs}
Our first goal in building an observation model is to identify and understand systematic effects in large-array bolometer surveys, relevant to cluster detection and cosmology. 
The second is to avoid to run a full Monte-Carlo chain to generate source catalogue observed by SZ-survey, a time consuming step in an analysis software that limits the number of possible iterations in partice. This is a strong assumption, that we checked up to the precision of the statistical uncertainties of upcoming surveys.

\subsection{Observation model ingredients}
Semi-analytic LSS model provides the expected number $N_{clusTh}$ of cluster of flux above a chosen threshold $Y_{Thres}$ and the cluster probability density function pdf($Y_{th}$, $z_{th}$). The observation model includes: 

-- A 2D selection function  Sel($Y_{th}$, $z_{th}$) giving the probability of a cluster of flux $Y_{th}$ and redshift $z_{th}$ to be detected. This 2D function is calibrated on simulation as described at paragraph \ref{CMonteCarl}. The expected number of true detected cluster can then be computed by integration of  
\beq \label{NbClusObs}
N_{obsTh} = \int N_{clusTh} . pdf(Y_{th},z_{th}) . Sel(Y_{th}, z_{th}) \quad dY_{th} dz_{th}
\eeq	

 -- Detection count variance model. In order to take into account the excess on true cluster counts described above, we introduce the pdf of observing $N_{obs}$ clusters, given $N_{obsTh}$,  pdf($N_{obs}$ $|$ $N_{obsTh}$). 
 We simply assume that it follows a gaussian law, with a width  $\sigma = A^{clus} \sqrt{N_{obsTh}}$, with $A^{clus}$ fitted to the value 3.0, as explained at paragraph \ref{SrcCounts}.
 
-- A photometric model pdf($Y_{obs}$, $z_{obs}$ $|$ $Y_{th}$, $z_{th}$) that provides the pdf of observing $Y_{obs}$ and $z_{obs}$ given $Y_{th}$ and $z_{th}$. 

-- A contamination model: we assume that contamination are driven by the foreground models and that these do not depend on cosmological parameters.  The pdf of the observed flux of contaminants pdf($Y_{Cont}$) is deduced from the Monte-Carlo simulations of paragraph \ref{VerminChap} after normalisation. The expectation of the contaminant counts  and its variance are taken to be constant and fitted from the distribution shown in figure \ref{ModelNDetection}.

-- An error model on the redshift determinations, when such a complementary measurement is available: pdf($z_{obs}$ $|$ $z_{th}$).

%
\begin{figure}[htp]
	\begin{minipage}[l]{.3\linewidth}
	\centering
	\includegraphics[width=5.5cm]{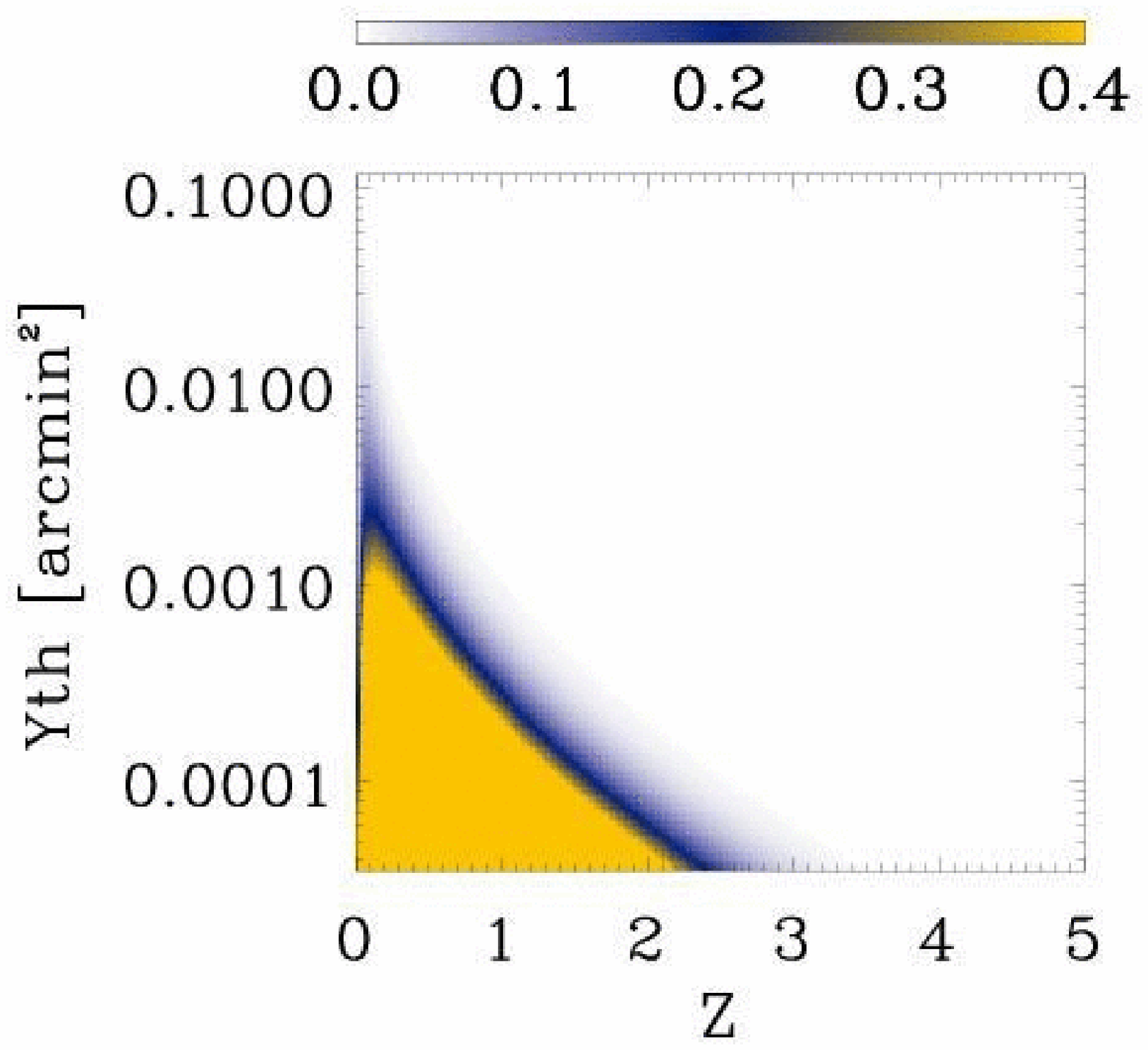}
	\end{minipage} 
\hfill
	\begin{minipage}[c]{.3\linewidth}
	\centering
	\includegraphics[width=5.5cm]{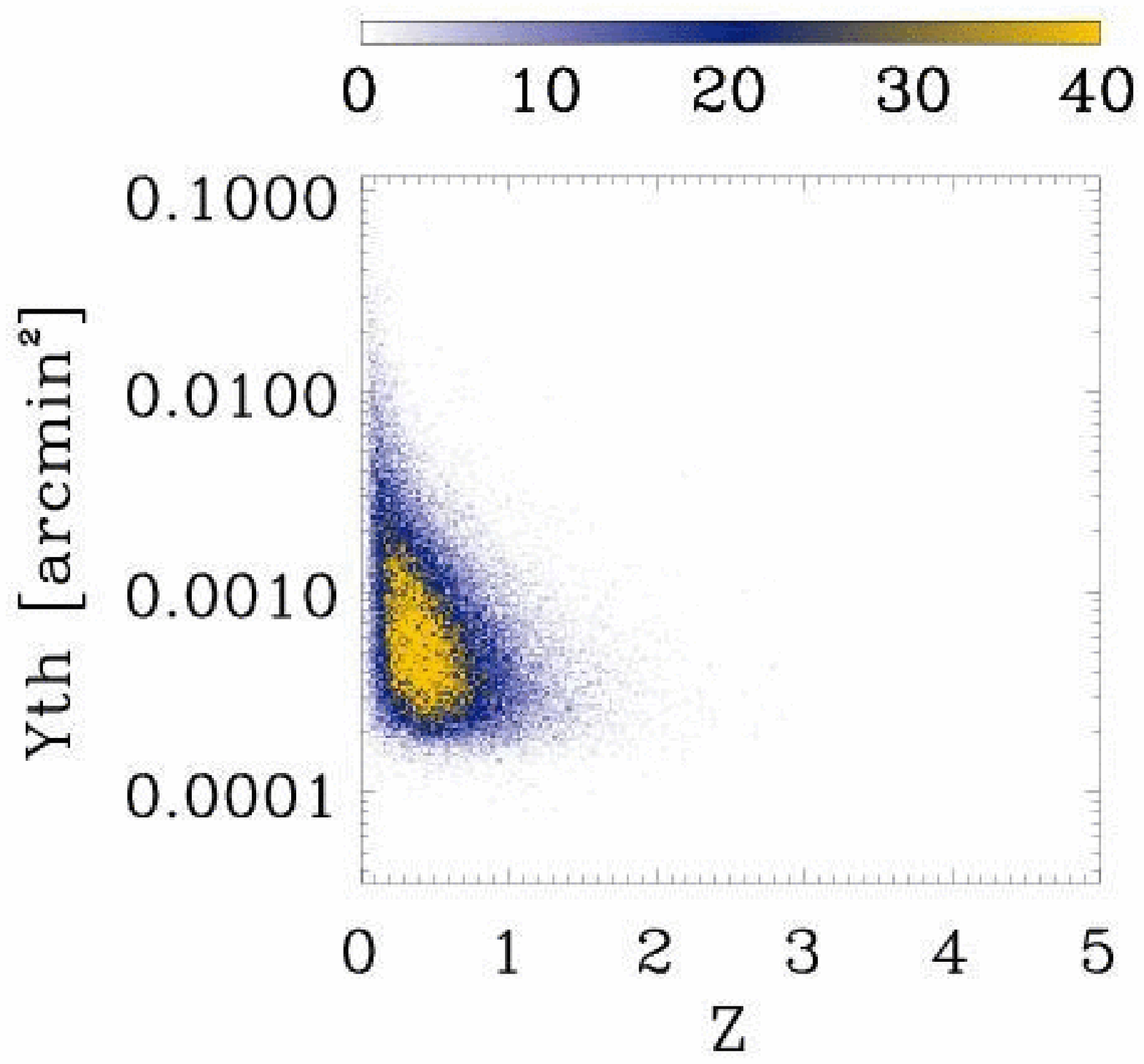}
	\end{minipage} 
\hfill
	\begin{minipage}[r]{.3\linewidth}
	\centering
	\includegraphics[width=5.5cm]{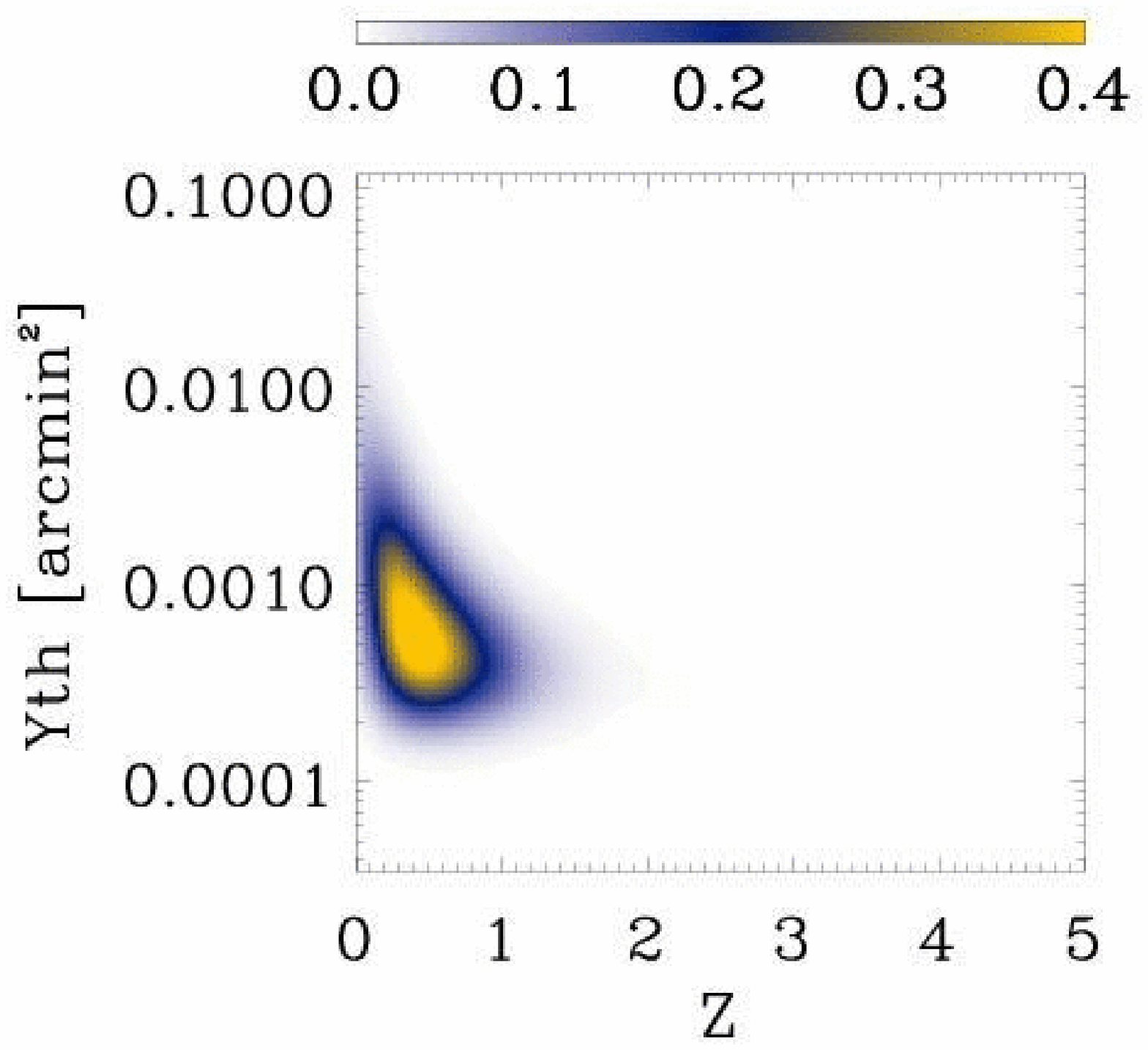}
	\end{minipage} 
	\caption[Observations'  model as seen by the eyes]
{\textbf{Left:} The cluster distribution generated by simulations, $\frac{dN^{amas}}{dz_{th} dY_{th}} (z_{th},Y_{th})$, and the observed cluster distribution, $\frac{dN^{amas}_{obs}}{dz_{th} dY_{th}} (z_{th},Y_{th})$, in 100 cumulative Monte-Carlo simulations \textbf{(middle)}, and from the observation model \textbf{(right)}. The axes are the integrated flux $Y$ in $\unit arcmin^{2}$ versus redshift.}
\label{ModelStatOlimpo}
\end{figure}

\subsection{Simplifying assumptions}
The above components of the observation model have been derived from the simulations, given instruments parameters and for the concordance cosmological model. They have been shown to be very sensitive to experiment properties such as noise level, number of observation bands, etc. When constraining cosmological parameters, these experimental effects are expected to be under control. 
On the other hand, we assume that the observation model is \emph{not sensitive}  to the cosmological parameters, when these are \emph{reasonably} close to our reference model. This assumption is strong and not obvious, since in large-array bolometers survey, the contribution of source confusion to the photometric noise may not be negligible. We checked the validity of this assumption, all other parameters being kept constant, by changing the cluster map density  by a factor 1.5 and 0.75.  Both recovered observation models were compatible with the above model, except for a small increase in the width of the photometry curve (paragraph \ref{PhotErrorsChap}) in the large density option.  We assumed this to be acceptable since, would such a dramatic discrepancy of cluster density be observed, we would recalibrate our observation model on representative simulations. 
 
\begin{figure}[htp]
\centering
\includegraphics[width=8cm]{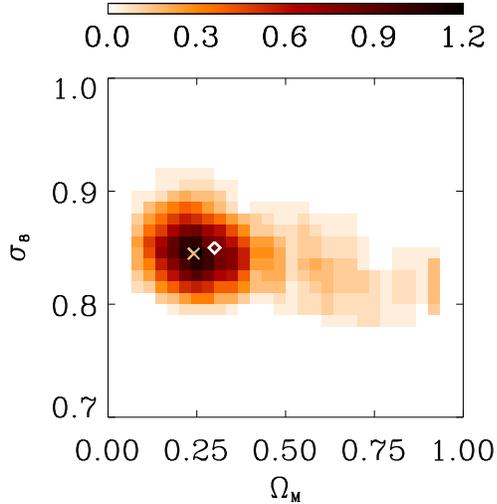}
\caption[Observations'  Model Cosmo Params Tests.]
{Probability density of the recovered cosmological parameters $\sigma_8$ and $\Omega_M$, for 100 Monte-Carlo simulations and using the observation model. Diamond is the model used at the input of simulations. Cross, is the maximun of occurrence of reconstructed parameters.}
\label{CheckModStatObs}
\end{figure}

Thus given a cosmological model and observations' model, we derive a set probability density function describing our observations: pdf($N_{Obs}$), pdf($Y_{obs}$, $z_{obs}$), pdf($Y_{Cont}$), pdf($N_{Cont}$). Figure \ref{ModelStatOlimpo} shows the distributions of flux and redshifts of detected clusters, generated by full Monte-Carlo and from our observation model. Those are remarkably similar, thus confirming the validity of our observation model. 

We also tested whether the use of the observation model would bias the cosmological parameter estimation. For this purpose, we computed 100 full Monte-Carlo source catalogues. For each catalogue, we computed the cosmological parameters using our observations' model. Figure \ref{CheckModStatObs} show the surface density of two of the fitted parameters $\sigma_8$ and $\Omega_M$ mostly relevant for this study. We observe that the input cosmological parameters are well within the 68\% CL contour of the distribution. Thus, the bias induced by the observation model is small compared to the statistical error of the observations.

We conclude that the use of an observation model is legitimate given the accuracy of upcoming experiments. This observation model will be improved: taking into account non gaussian tails in our photometry model is the main improvement foreseen. In the following, all source catalogues have been generated using the observation model.

\section{Cosmological implications} \label{CosmoChap}
The mains physics goals of large SZ-cluster surveys are to learn more about cluster gas physics, large scale structure, and cosmology. These physical models are parametrised, and  involve assumptions that can be tested by future SZ cluster surveys. In the following, we first present statistical tools and results testing the compatibility of our data, with a parametrised model family in paragraph \emph{hypothesis tests}. Then we show how SZ cluster data can constraint the mass temperature normalisation factor $T_*$, using a classical \emph{parameter estimation} method.
Assuming then $T_*$ known, we explore the potential of SZ cluster survey, for constraining cosmological parameters $\Omega_M$ and $\sigma_8$.
We conclude by showing the effect on cosmological parameters of oversimplifying features of the observations' model.
In the following we assumed we have available a catalogue of observed sources corresponding to a nominal Olimpo scientific flight: 500 sources observed over 300 square degrees.

\subsection{Extended likelihood}
The tool for all the following statistical tests is the so called \emph{extended likelihood} of the cosmological parameters $\vec{C}$, given the experiment exp: $L(\vec{C} | exp)$. 
\beq
\label{Vraisemblance}
L (\vec{C} | exp) = \frac{dP}{dN^{Sour}} (N^{Sour}; \vec{C}) \, \prod_{i = 1}^{N^{Clus}} pdf( z^{Clus}_i, Y^{Clus}_i;\vec{C} ) \, \prod_{i = 1}^{N^{Sour}_{1}} pdf( Y^{Sour}_i;\vec{C} )
\eeq
with $N^{Sour} = N^{Clus} + N^{Sour}_{1}$ \\
The likelihood incoporates three kinds of information available in the data.The first factor is the probability of observing $N^{Sour}$ sources given the cosmological parameters $\vec{C}$.
The second factor is the probability of observing a cluster with a flux $Y$ and at redshift $z$ (using follow-up observations). We assume that the follow-up observation established whether the source is a cluster of galaxies, or a false detection. In the latter case, this source is excluded from the likelihood, except from the first factor. The third factor is the probability of observing a source of flux $Y$, when no follow-up observations were available. In this case, we do not know whether this source is a SZ-cluster or a false detection. Our observation model (paragraph \ref{StatModelObs}) provides these three factors in the likelihood, either directly or after integration and normalisation of the distributions.

\begin{figure}[htp]
\begin{minipage}[l]{.5\linewidth}
\centering
\includegraphics[width=8cm]{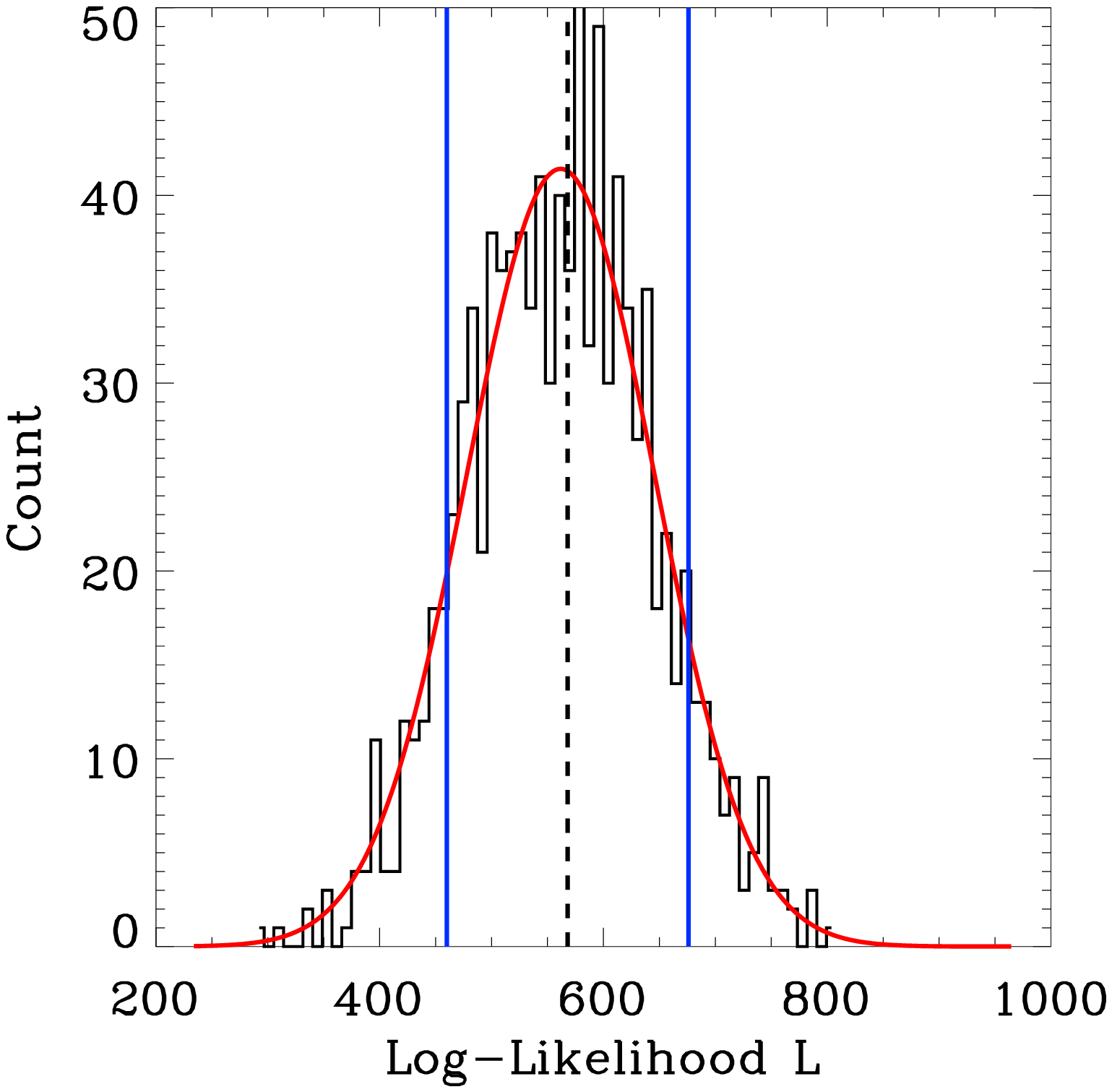}
\end{minipage} 
\hfill
\begin{minipage}[r]{.5\linewidth}
\centering
\includegraphics[width=8cm]{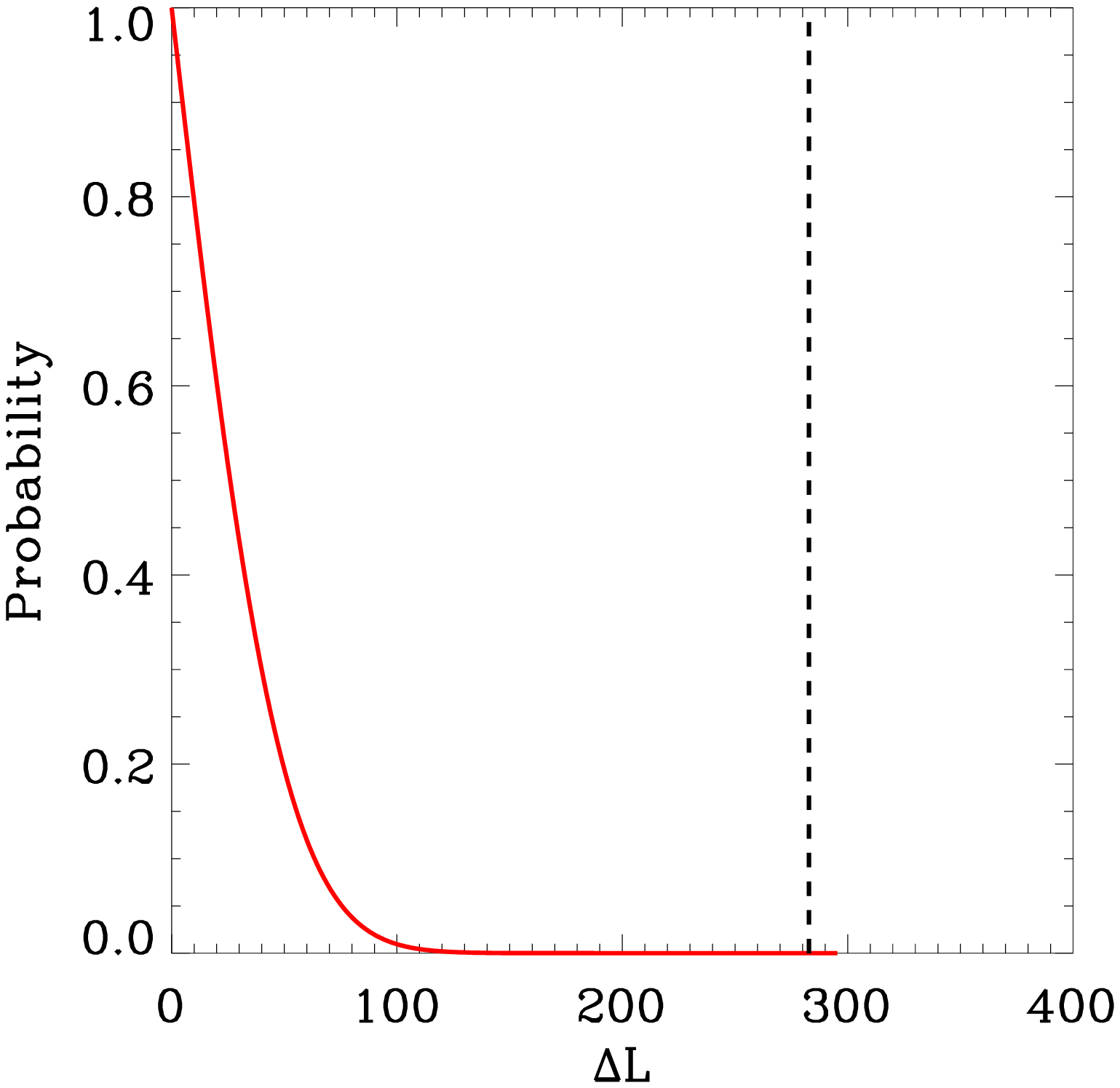}
\end{minipage} 
\caption[Test d'hypothse]
{\textbf{Left:} histogram of log-likelihood $L$ (black) for N Monte-Carlo catalogue of a Press-Schechter cosmological model. The peak is fitted by a gaussian law (red line), with mean $L_{mean}$. \\
\textbf{Right:} red line is the probability versus $\Delta L= L-L_{mean}$ of observing a Press-Schechter based catalogue with $\Delta L$. Vertical dashed line is the $\Delta L$ computed for a catalogue  generated from a Sheth and Tormen model. The probability of compatibility is lower than $10^{-5}$. }
\label{HypothesisTest}
\end{figure}

\subsection{Hypothesis tests}
The  question we wish to answer \emph{before} constraining parameters models is whether there
exist a parametrised model which is compatible with our data. To settle this issue, we use an hypothesis test method.  For a cosmological model, we generate by Monte-Carlo a large number of observed source catalogues, compute their likelihood for the given cosmological model and build an histogram of the likelihood (see figure \ref{HypothesisTest}).
The normalisation of the integral of the histogram provides the probability curve of an observed catalogue  to be compatible with the cosmological model. The use of the statistical observation model speeds up this work dramatically. Figure \ref{HypothesisTest} \emph{left}, shows the histogram of the likelihoods computed assuming the concordance cosmological model and a Press and Schechter mass function \citep{PS1974} for the clusters.
The black line shows the likelihood value computed for a source catalogue computed with the same cosmology, but with the mass function of Sheth and Tormen \citep{ShethMoTormen2001}. The probability of compatibility is lower than $10^{-5}$. The Press and Schechter hypothesis is thus rejected by the data.
In practise, cosmological parameters are free parameters and we often obtain a compatibility valley for our parameters with our ``observed" data. Other sources of constraints on cosmological parameters (such as CMB anisotropies) will allow us to select the relevant cosmological models and mass function.

\begin{figure}[htp]
\centering
\includegraphics[width=8cm]{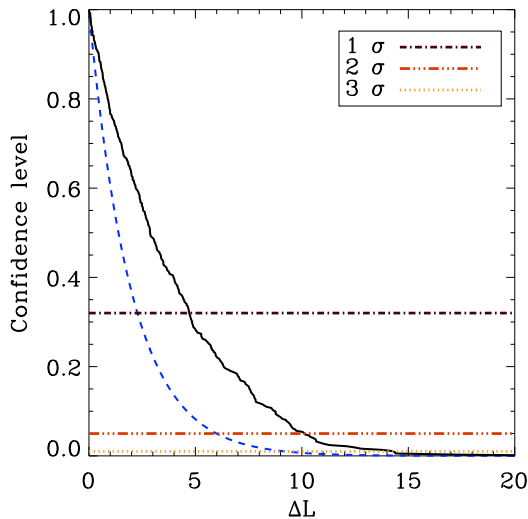}
\caption[Contraindre correctement les paramtres cosmos]
{\textbf{Left:} histogram of difference of log-likelihood $\Delta$ (black) for N Monte-Carlo catalogues of the best cosmological model $\vec{C_{min}}$ according to our data set. Dashed blue line is the $\chi^2$ law, expected for Gaussian distributions with 2 degrees of freedom. 68 \% and 95\% confidence levels are shown as dashed black and red lines. The $\chi^2$ approximation is very optimistic.} 
\label{NormCumDistrib}
\end{figure}

\subsection{Parameter Estimation}
Once we have selected a parametrised model compatible with out data, the next question is to
estimate a set of best cosmological parameters, in agreement with data and to compute the associated errors (or confidence levels).
%
For this purpose, we minimise the function $-lnL(\vec{C} |  \textbf{D})$ over $\vec{C}$, vector of the model parameters, to find the best model according to our data $\vec{C_{min}}$ and his likelihood $L_{min}$.
Then, we generate, according to $\vec{C_{min}}$, many source catalogues   \textbf{$C_i$},
minimise the likelihood to find the best Model $\vec{C_i}$ matching \textbf{$C_i$}, and build the histogram of 
\beq
\Delta _i = -2 \left[ ln L( \textbf{D} | \vec{C_i} ) -  ln( L_{min} ) \right] \\
\eeq
and the map of $\Delta _i$ at position $\vec{C_i}$. 
Computing the normalised cumulative distribution of the variable $\Delta$ (figure \ref{NormCumDistrib}) allows us to compute $\Delta _i $ values of the $68\%, 95\%$ and $99\%$ confidence level limits to be used to draw contours on the model map. Would the pdf be gaussian distributed, $\Delta _i$ distribution would follow a $\chi^2$ law. We notice that cluster density versus redshift and flux are \emph{not gaussian}.

\begin{figure}[htp]
\centering
\includegraphics[width=8cm]{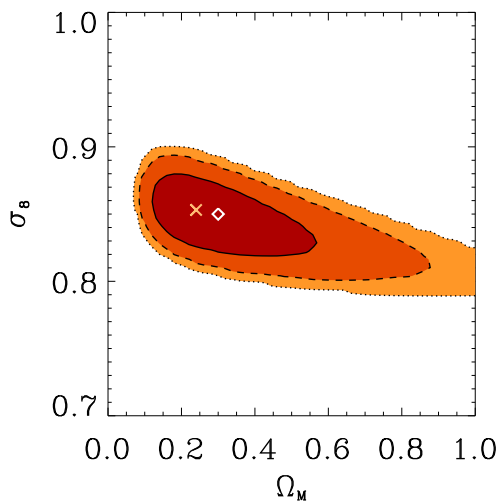} 
\caption[]
{Expected constraints on $\sigma_8$ and $\Omega_M$ from an Olimpo scientific flight, with full spectroscopic follow-up of the sources. All other cosmological parameters, have been set to the values in table \ref{CosModParam}. These constraints are severely degraded due by the excess count variance observed in the Monte-Carlo simulations.
}
\label{CosmoCons}
\end{figure}

\subsection{Constraints on cosmological parameters for the Olimpo SZ-cluster surveys}
We now use the above tools to constraint the cosmological parameters. In the following, we will assume a $\Lambda$CDM cosmological model with parameter list of table \ref{CosModParam}. The most important parameters for large scale structure formation and SZ-clusters are $\sigma_8$ and $\Omega_M$ as well as $T_*$, the normalisation of the mass to temperature scale relations \citep{Pierpaoli2001} in cluster formation models. We plot in figure \ref{CosmoCons} the expected constraints on $\sigma_8$ and $\Omega_M$ from SZ-cluster observations, assuming all other cosmological parameters known, at their simulated values. We assumed that follow-up observations provided redshifts for \emph{all} the sources. This is the best constraint achievable, according to our observation model.


We observed that these constraints are rather pessimistic, given the ambition of the SZ-cluster projects. This is due to the excess width in the count variance, observed in our Monte-Carlo simulations, equivalent to a statistic loss of a factor 10.

\begin{figure}[htp]
\begin{minipage}[l]{.5\linewidth}
\centering
\includegraphics[width=8cm]{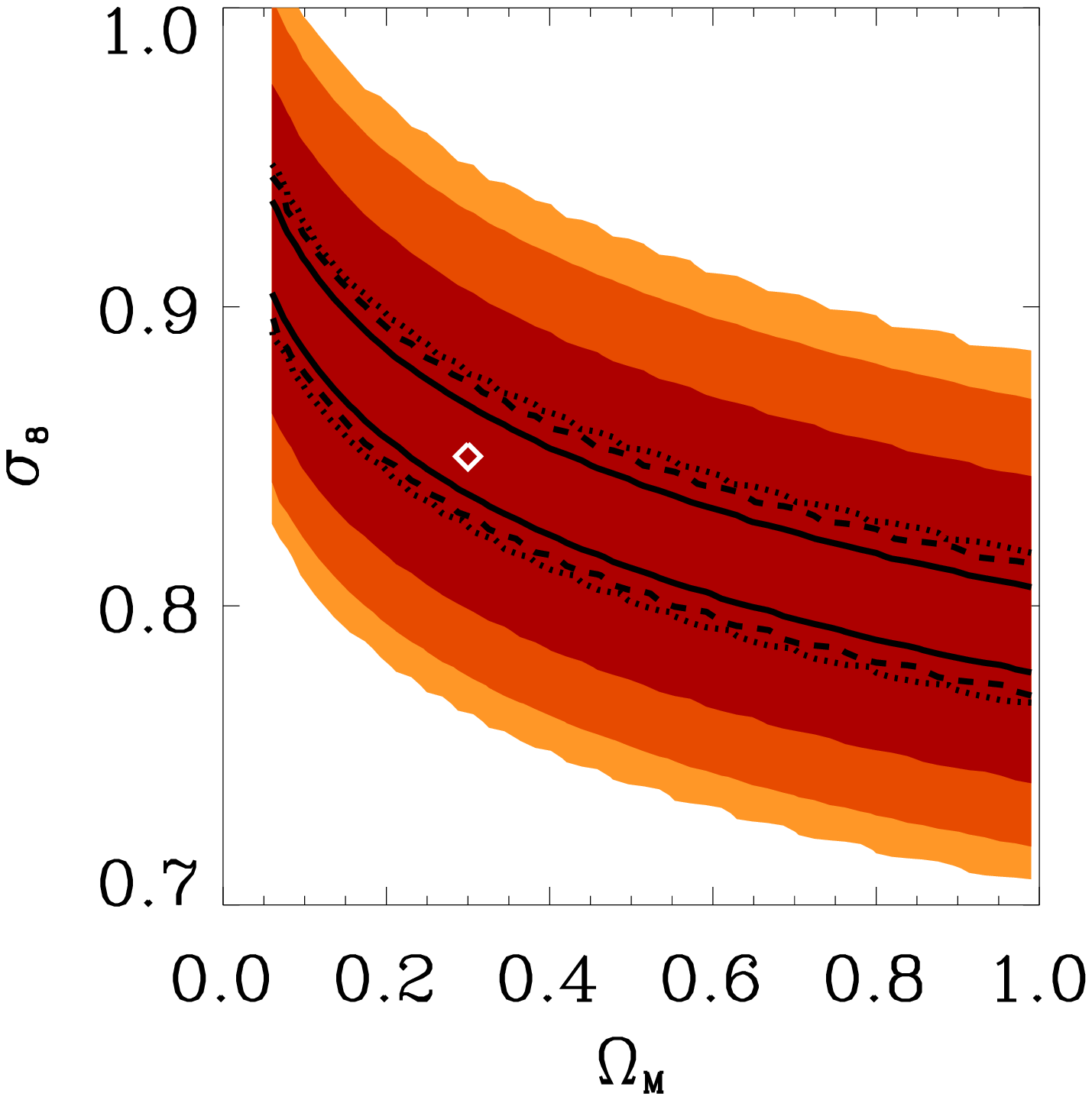}
\end{minipage} 
\hfill
\begin{minipage}[r]{.5\linewidth}
\centering
\includegraphics[width=8cm]{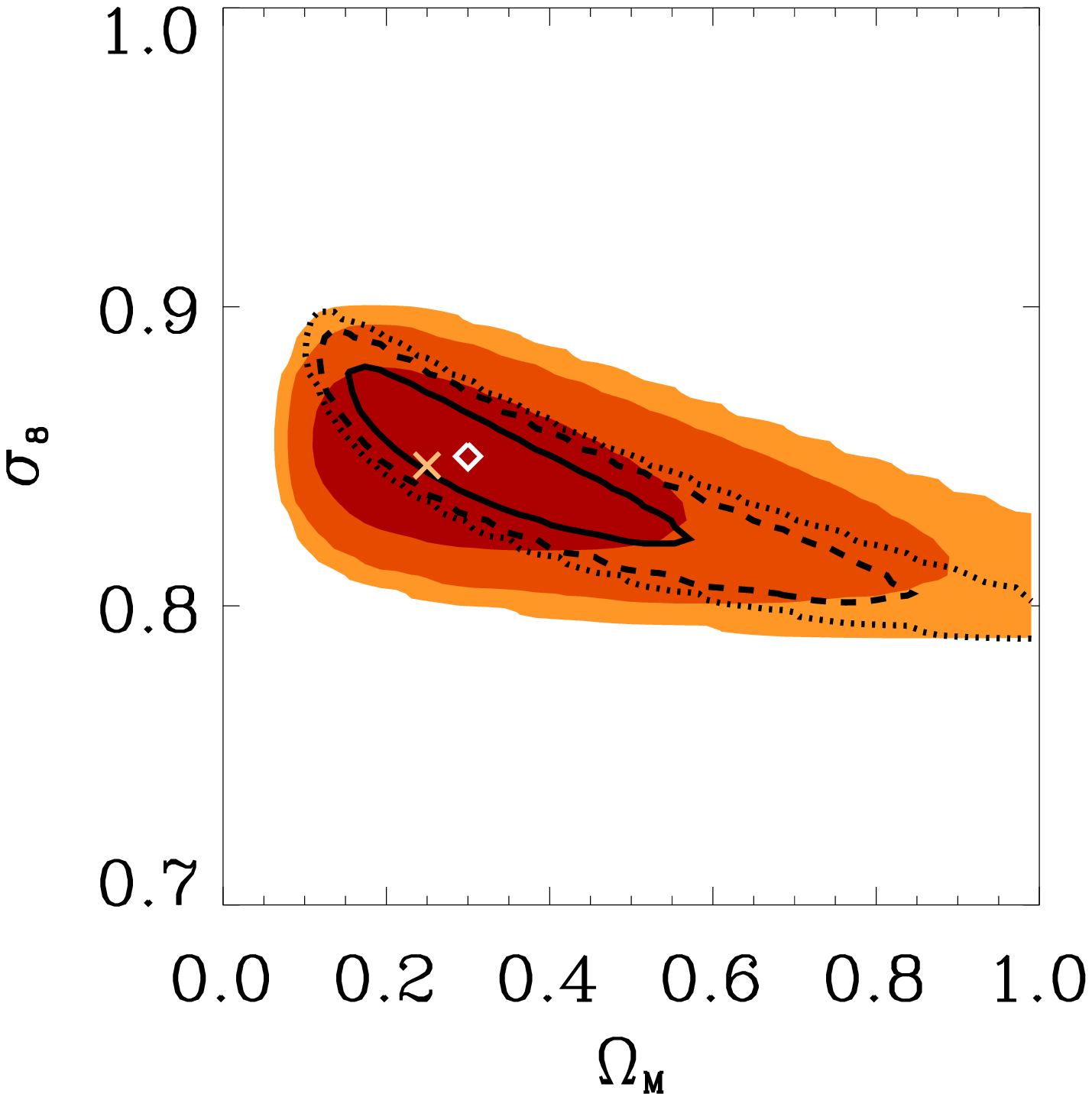}
\end{minipage} 
\caption[]
{Degradations of constraint due to the cluster count variance. 
 \textbf{Left:}  Colored  are the $\Omega_M$ vs $\sigma_8$ Confidence Level contours computed using the observed Monte-Carlo count variance, all other cosmological parameters set to their simulated values.  Only the information on SZ-cluster count has been used in this figure, \emph{no redshift}. Dashed are the same CL constraints, assuming a Poisson-law field to field cluster count variation.\\
\textbf{Right:}  $\Omega_M$ $\sigma_8$ Confidence Level contours assuming 100\% follow-up for cluster redshifts, assuming a Poisson-law field to field cluster count variation, all other cosmological parameters set to simulated values. Colored contours are a copy of figure \ref{CosmoCons}}.
\label{CountVariance}
\end{figure}

\subsubsection{Cluster count variance} \label{CountVarChap}
As shown at paragraph \ref{SrcCounts} the simulated source count variance is larger than a Poisson's distribution of same expectation that we would assume from field to field variance. If we now forget  the simulation results, the cluster count carries richer cosmological information which is quantified in the first term of the likelihood. Figure \ref{CountVariance} shows the constraints computed with this optimistic assumption. We have not been able to decide yet where does the count variance comes from. First suspect is the cluster detection software, with an out of control feature. Second is the confusion due to the few large clusters, that mask smaller cluster and thus induces lower statistic processes in the count variance. Third suspect is the effect of one of the foreground. Therefore cluster detection algorithms for cosmology, \citep{Herranz2002a, Pierpa05, MelinThese, PiresJuin2005}, should be evaluated, on their efficiency, on the purity of the recovered source catalogue, but definitely \emph{on the source count variance} at the output of the chain.
We are working on this issue. Solving/minimising this effect is now our first priority. 

\begin{figure}[htp]
\begin{minipage}[l]{.50\linewidth}
\centering	
\includegraphics[width=8cm]{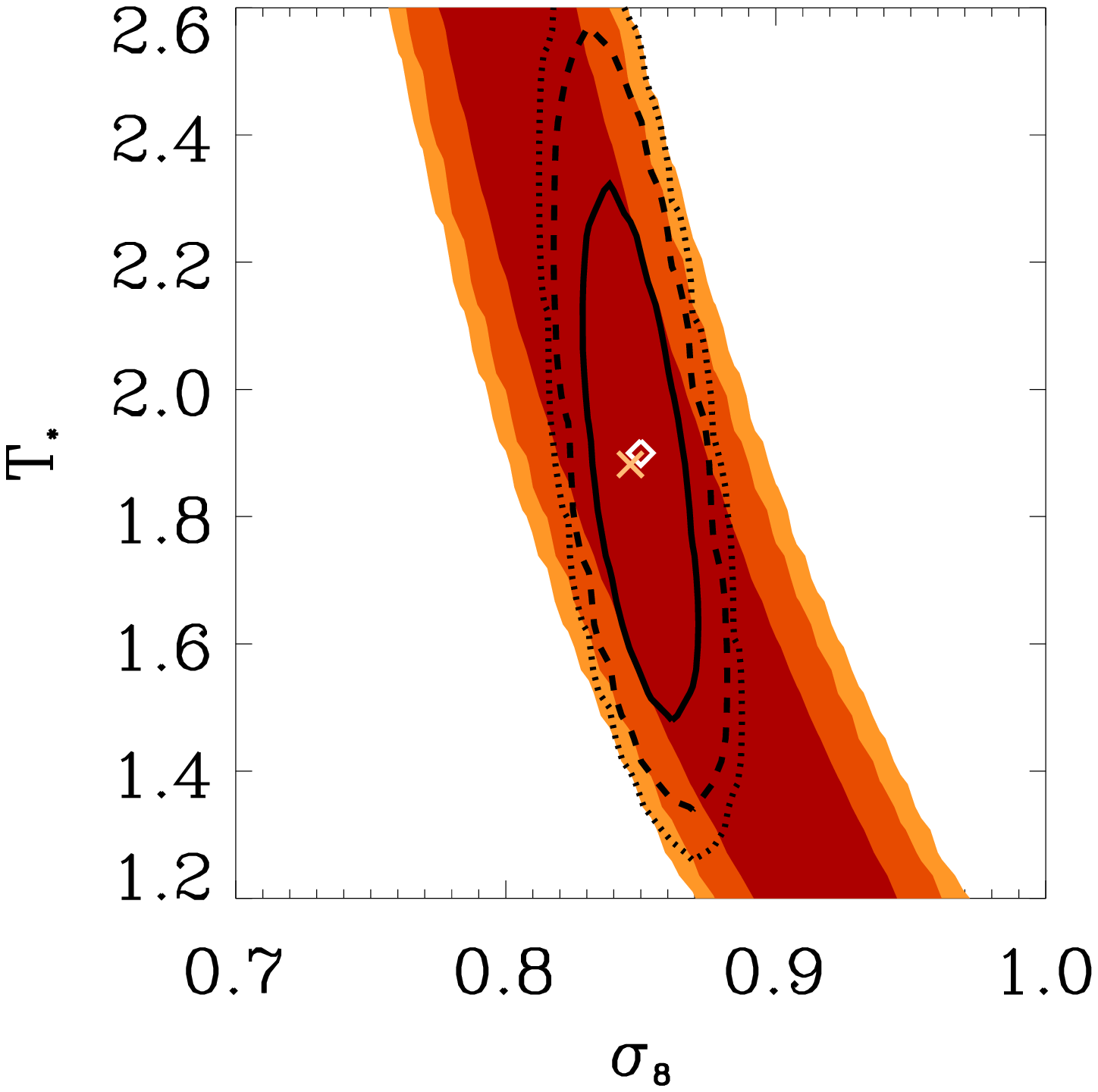}
\end{minipage} 
\hfill
\begin{minipage}[r]{.50\linewidth}
\centering
\includegraphics[width=8cm]{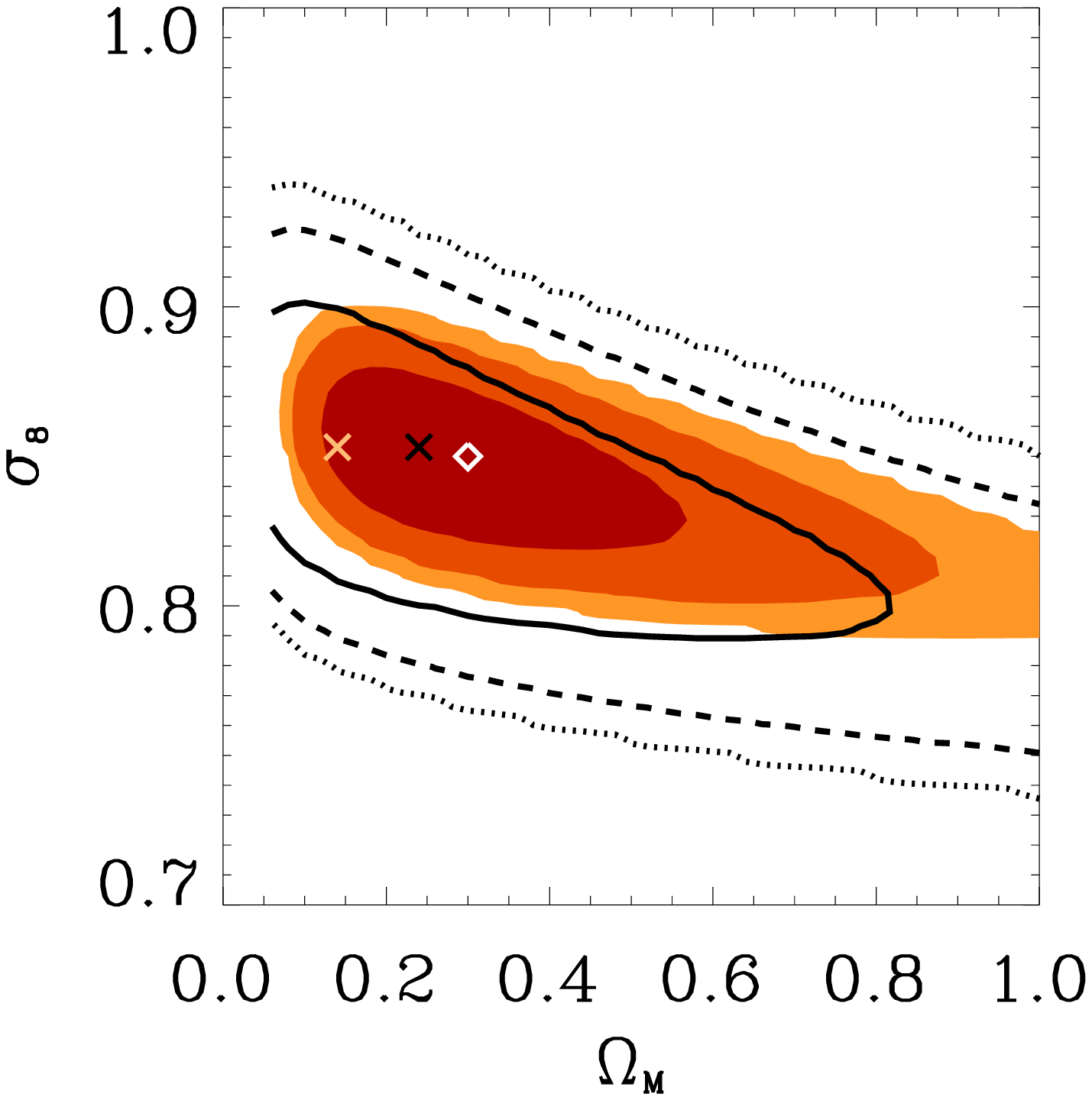}
\end{minipage} 
\caption[]
{\textbf{Left:} confidence level map, on $\sigma_8$ and $T_*$, marginalised on  $\Omega_{M}$. Colors are computed from SZ-cluster data only, dashed lines uses WMAP and CFHTLS weak-shear Fisher matrix constraints, no systematic effect are included.\\
\textbf{Right:} Lines are the constraints on cosmological parameters if we keep only the largest flux clusters. All other cosmological parameters, have been set. Diamond is the generated concordance model. White cross is the reconstructed model. Colors delimit the references CL contour. Lower statistic induces heavy loss in the constraint accuracy.}
\label{DegLatePhys}
\end{figure}

\subsubsection{Degeneracy with late cluster physics}
Galactic physics event heats the extragalactic cluster gas, and thus contribute to the SZ-cluster signal in addition to the gravitational potential and virialisation. Late cluster gas heating mechanisms are not well known yet. Their contribution to gas heating is commonly parametrised in the mass-temperature relation as a normalisation factor, $T_*$. Figure \ref{DegLatePhys} shows the CL contours placed on $T_*$ and  $\sigma_8$ marginalised on $\Omega_M$. We observe that with the input of WMAP and CFHT-LS weak-shear forecast, the residual correlation between  $T_*$ and $\sigma_8$ is small. In addition on going X-Rays surveys from XMM satellite should provide a lot of information on cluster gas physics and allow precise determination of $T_*$. Thus in the following we set $T_*$ to the value $1.9$.

\subsubsection{Restriction to high-mass clusters}
Large clusters involve hundred of galaxies. Their gravitational potential is stronger than in smaller clusters and non-gravitational physics is less important than in low mass systems.
As a result, their mass to temperature scaling law is expected to show a smaller dispersion. Thus one can foresee that heavy clusters will be statistically better modelled, and that constraints based on massive cluster observations will be reliable.
It is thus instructive to study the cosmological constraints which can be derived from a
sample restricted to high-mass clusters. Figure \ref{DegLatePhys} shows the confidence level map computed from a catalogue, when we select clusters of flux larger than $7.4 \times 10^{-4} \unit arcmin^2$. The CL contour are significantly enlarged compared to the reference contour drawn in black, because of the much smaller statistic. This is a strong motivation for theorists to understand and build a model of low-mass clusters.

\begin{figure}[htp]
\begin{minipage}[l]{.5\linewidth}
\centering
\includegraphics[width=8cm]{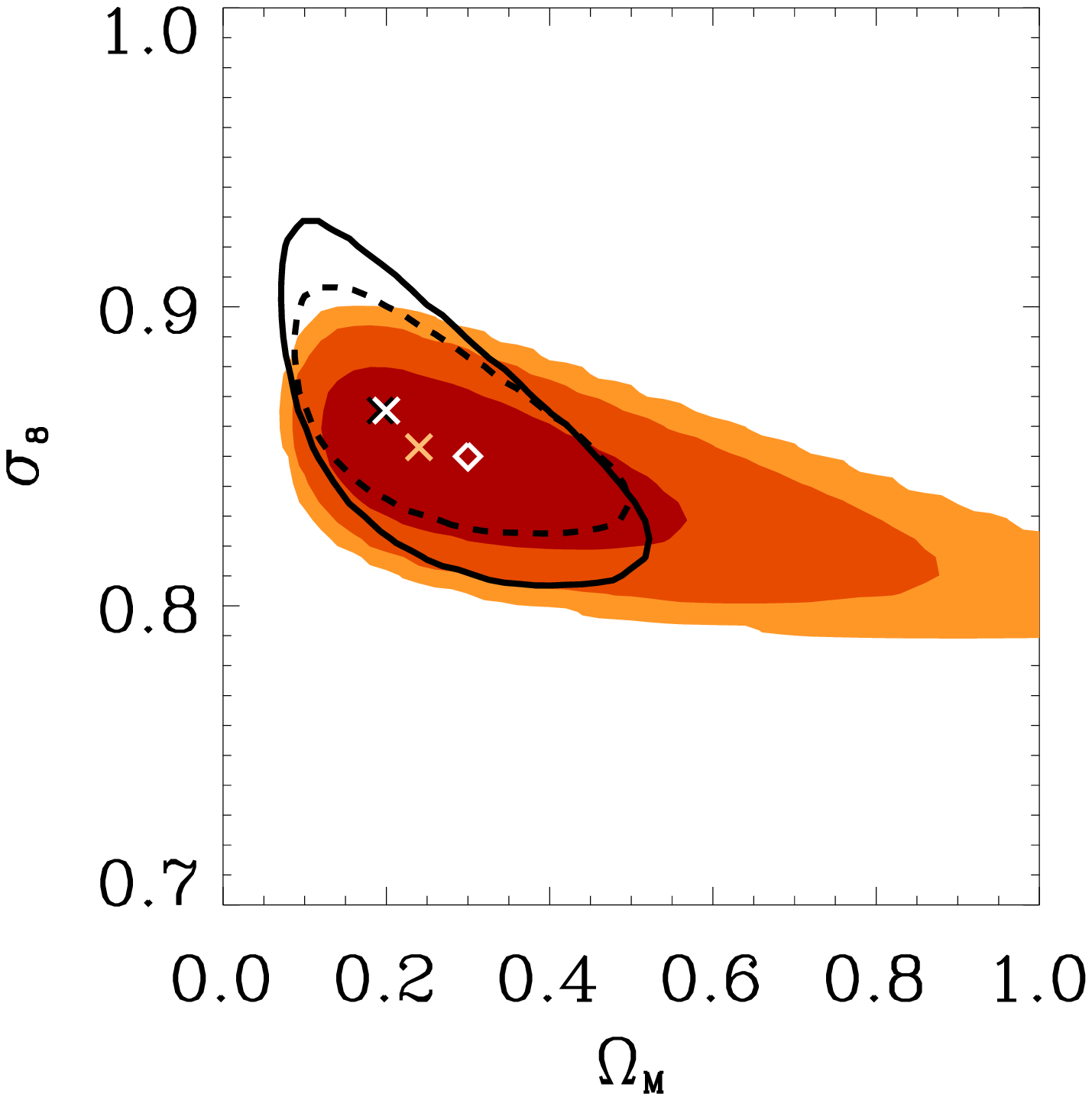}
\end{minipage} 
\hfill
\begin{minipage}[r]{.5\linewidth}
\centering
\includegraphics[width=8cm]{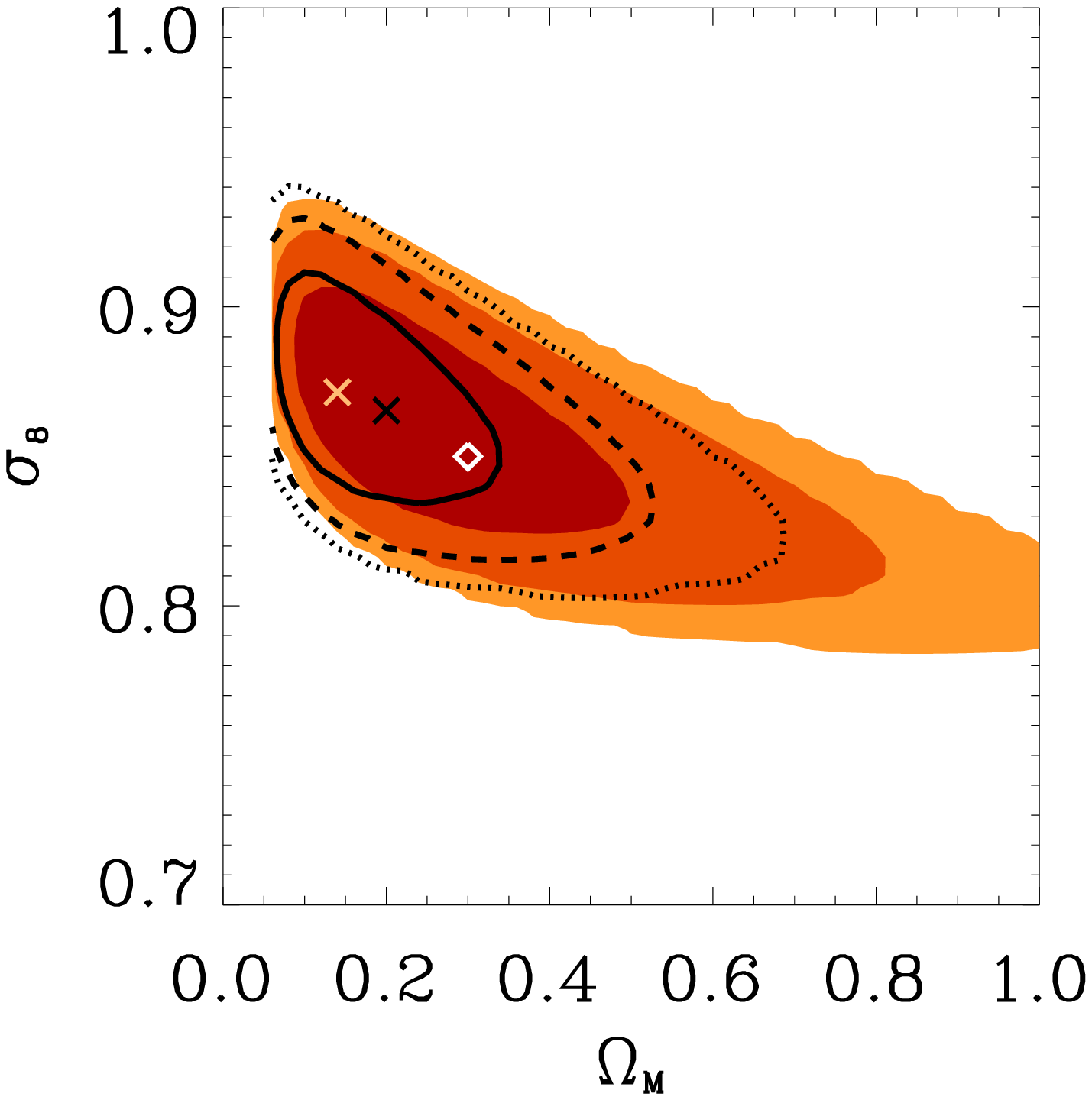}
\end{minipage} 
\caption[]
{\textbf{Left:} impact on cosmological constraints, due to an incomplete redshift follow-up of cluster candidates. Black line is the 68 \% CL contour assuming a 10\% coverage follow-up. Dashed line assumes 50\% coverage follow-up. Colored contours are a copy of figure \ref{CosmoCons}.\\
\textbf{Right:} Lines show the systematic shift in the CL contour induced by neglecting contaminants in the recovered source catalogue. This plot was generated assuming that 50\% of the sources have been observed in follow-up for redshift. Colors stand for contours computed with the same dataset, but taking into account contaminations.
}
\label{SysteEffects}
\end{figure}

\subsubsection{Incomplete spectroscopic follow-up}
Large-array bolometer surveys will provide large cluster catalogues, including flux and positions and shapes for resolved clusters, but have to rely on follow-up experiments for redshift measurements. Figure \ref{SysteEffects} show the impact of partial redshift coverage, assuming 10\% or 50\% random coverage of clusters candidates. Remaining catalogue  contaminations have been properly taken into account. We note that no significant bias on the CL contours is seen, but that the statistics are degraded.
This shows that follow-up observations will be very important for the accuracy of the physics output of large SZ-Cluster surveys.

\subsubsection{Neglecting contamination}
Assuming now that the redshift followup of the observations will be incomplete (as is very likely to be the case in practice), we quantify now the effect of neglecting contamination in the recovered catalogue. We assume in the paragraph that 50\% of the sources have a redshift. Technically, this means using in the first factor of the likelihood the pdf count of true detected cluster and in the third factor, the integral over $z_{obs}$ 
of $\frac{dP}{dz_{obs} dY_{obs}} (z^{Clus}_i, Y^{Clus}_i;\vec{C})$. Figure \ref{SysteEffects} shows the results. We see that since we assumed that false detections are clusters, the reconstructed parameters are biased toward large $\sigma_8$, since contamination produce a spurious enhancement of the  cluster distribution at low surface brightnesses. The reduced size of the CL contour is a secondary effect of the low $\Omega_M$ fitted value.

\section{Conclusion}
In this paper, we have explored, in details, the potential and limitations of upcoming wide-field SZ surveys. We used a full simulation pipeline, going from the cosmological models to recovered cluster catalogues and constraints on cosmological parameters. We showed that the selection function and purity of the recovered catalogues are more complex than is usually assumed. We quantified the foreseen selection function, photometry and contamination of the upcoming Olimpo project.\\
We observed a cluster count variance larger than Poisson's width, when running several Monte-Carlo simulations. This effect worsens the cosmological constraints. We did not identify yet the origin of this excess width, but reducing it is our first development priority in order to achieve the full potential of SZ-surveys for cosmology. \\ 
We presented methods to statistically model the observations, select parametrised models compatible with data, and then constraint models parameters. We showed that, using SZ Cluster data, combined with WMAP and expected CFHTLS weak-shear forecast data,  little correlation is seen between efficiently the mass-temperature normalisation factor $T_*$ and $\sigma_{8}$. Complementary X-Rays observations will be necessary to put tighter constraints on $T_*$, but on the other hand, we only need moderate precision on $T_*$, to achieve good knowledge on $\sigma_{8}$. We finally exposed the impact on cosmological parameters of systematics in observations or interpretation of our data. \\
This paper does not use the 2-point correlation of SZ Cluster to constraint cosmology \citep{MeiBartlett2003}. This incorporation of the 2-point correlation function in our simulation pipeline and its
impact on cosmological parameter is left to future work.

\section{Acknowledgements}
We hereby acknowledge many scientific and algorithmic discussions with J. Rich, J.-P. Pansart, C. Magneville, R. Teyssier (CEA Saclay/DAPNIA) and  J.-B. Melin and  J. G. Bartlett (Univ. Paris 7, APC). Special thanks to P. Lutz and J. Bouchez (CEA Saclay/DAPNIA) for their help with statistic methods.

\section{Annex: selection function data} \label{SelFuncAnnex}
The following table \ref{SelFuncTab} samples values of the selection function as a function of mass and Compton flux.

\begin{table}[hhh]
\centering
\begin{tabular}{|c||c|c|c|c|}
\hline
Redshift & 90\% efficiency flux [$\unit arcmin^2$] & 90\% efficiency Mass [$\unit M_{sun}$] \\
\hline
\hline
0.1 & $2.5\;10^{-3}$ & $2.9\;10^{14}$ \\
\hline
0.2 & $1.5\;10^{-3}$ & $3.8\;10^{14}$ \\
\hline
0.3 & $1.0\;10^{-3}$ & $4.0\;10^{14}$ \\
\hline
0.4 & $7.7\;10^{-4}$ & $4.5\;10^{14}$ \\
\hline
0.5 & $6.5\;10^{-4}$ & $5.0\;10^{14}$ \\
\hline
0.7 & $5.6\;10^{-4}$ & $5.1\;10^{14}$ \\
\hline
1.0 & $4.9\;10^{-4}$ & $5.1\;10^{14}$ \\
\hline
2.0 & $4.2\;10^{-4}$ & $4.2\;10^{14}$ \\
\hline
3.0 & $4.2\;10^{-4}$ & $3.2\;10^{14}$ \\
\hline
4.0 & $4.2\;10^{-4}$ & $2.2\;10^{14}$ \\
\hline
5.0 & $4.2\;10^{-4}$ & $2.0\;10^{14}$ \\
\hline
\end{tabular}
\caption{Selection function versus redshift. Value entered correspond to 90\% efficiency.}
\label{SelFuncTab}
\end{table}
%


\bibliographystyle{aa}

\bibliography{Biblio}

\end{document}